\documentclass[11pt,a4paper]{article}
\usepackage{jcappub}

\usepackage{epsfig,psfrag,graphics,verbatim}
\usepackage{graphicx}

\usepackage{dcolumn}
\usepackage{bm}
\usepackage{color}
\usepackage{ulem}
\newcommand{\neut}{{\tilde{\chi}^0_1}}
\newcommand{\params}{{\mathbf \Theta}}

\newcommand{\be}{\begin{equation}}
\newcommand{\ee}{\end{equation}}
\newcommand{\sigmaSI}{\sigma_{\neut-p}^\text{SI}}
\newcommand{\sigmaSD}{\sigma_{\neut-p}^\text{SD}}

\newcommand{\BR}{BR}
\newcommand\RBtaunu{\frac{\BR(B_u \to \tau \nu)}{\BR(B_u \to \tau \nu)_{SM}}}
\newcommand\DeltaO{\Delta_{0-}}
\newcommand\RBDtaunuBDenu{\frac{\BR(B \to D \tau \nu)}{\BR(B \to D e \nu)}}
\newcommand\Rl{R_{l23}}

\newcommand\Dstaunu{\BR(D_s \to \tau \nu)}
\newcommand\Dsmunu{\BR(D_s\to \mu \nu)} 
\newcommand\Dmunu{\BR(D \to \mu \nu)} 
\newcommand\brbsmumu{\BR(\overline{B}_s\to\mu^+\mu^-)}

\begin{document}


\title{Updated global fits of the cMSSM including the latest LHC SUSY and Higgs searches and XENON100 data}

\author[a]{C. Strege}
\author[b]{G. Bertone}
\author[c,d]{D. G. Cerde\~{n}o}
\author[e,1]{M. Fornasa \note{MultiDark fellow}}
\author[f]{R. Ruiz de Austri}
\author[a]{R. Trotta}

\affiliation[a]{Astrophysics Group, Imperial College London, Blackett Laboratory, Prince Consort Road, London SW7 2AZ, UK}
\affiliation[b]{GRAPPA Institute, University of Amsterdam, Science Park 904, 1090 GL Amsterdam, Netherlands}
\affiliation[c]{Instituto de F\'{\i}sica Te\'orica UAM/CSIC, Universidad Aut\'onoma de Madrid, 28049 Madrid, Spain}
\affiliation[d]{Departamento de F\'{\i}sica Te\'orica, Universidad Aut\'onoma de Madrid, 28049 Madrid, Spain} 
\affiliation[e]{Instituto de Astrof\'{\i}sica de Andaluc\'{\i}a (CSIC), E-18008, Granada, Spain}
\affiliation[f]{Instituto de F\'isica Corpuscular, IFIC-UV/CSIC, Valencia, Spain} 

\abstract{
We present new global fits of the constrained Minimal Supersymmetric Standard Model (cMSSM), including LHC 1/fb integrated luminosity SUSY exclusion limits, recent LHC 5/fb constraints on the mass of the Higgs boson and XENON100 direct detection data. Our analysis fully takes into account astrophysical and hadronic uncertainties that enter the analysis when translating direct detection limits into constraints on the cMSSM parameter space. We provide results for both a Bayesian and a Frequentist statistical analysis. We find that LHC 2011 constraints in combination with XENON100 data can rule out a significant portion of the cMSSM parameter space. Our results further emphasise the complementarity of collider experiments and direct detection searches in constraining extensions of Standard Model physics. The LHC 2011 exclusion limit strongly impacts on low-mass regions of cMSSM parameter space, such as the stau co-annihilation region, while direct detection data can rule out regions of high SUSY masses, such as the Focus-Point region, which is unreachable for the LHC in the near future. We show that, in addition to XENON100 data, the experimental constraint on the anomalous magnetic moment of the muon plays a dominant role in disfavouring large scalar and gaugino masses. We find that, should the LHC 2011 excess hinting towards a Higgs boson at 126 GeV be confirmed, currently favoured regions of the cMSSM parameter space will be robustly ruled out from both a Bayesian and a profile likelihood statistical perspective.
}




\maketitle

\section{Introduction}
\label{secintro}

The CMS and ATLAS collaborations have recently published new results for the search for Supersymmetry (SUSY) \cite{arXiv:1111.4820,arXiv:1109.2352,arXiv:1109.6572}, based on proton-proton collisions with a center-of-mass energy $\sqrt{s} = 7$ TeV, and an integrated luminosity of 1 fb$^{-1}$, recorded during 2011. No significant excess above the Standard Model (SM) predictions was measured, so that new constraints on SUSY could be derived, in particular for what concerns the cMSSM (constrained Minimal Supersymmetric Standard Model, see Refs. \cite{Chamseddine:1982jx,CTP-TAMU-24-92,RAL-92-005,hep-ph/9311269,Kane:1993td} and references therein), a simple construction of minimal SUSY widely studied in the literature, which has become the {\em de facto} default model to evaluate SUSY searches and, often, prospects for dark matter searches \cite{Bertone:2010zz,Bergstrom:2000pn,Munoz:2003gx,Bertone:2004pz}. Additionally, an update on the search for the SM Higgs boson based on 5 fb$^{-1}$ of data has recently been announced by the ATLAS and CMS experiments \cite{Higgs_results}. In particular, the ATLAS experiment has reported a new 95$\%$ exclusion region constraining the allowed mass range for the SM Higgs boson to $115.5 - 131.0$ GeV, as well as a 2.3$\sigma$ excess at a mass of $126$ GeV. 


In Ref.~\cite{arXiv:1107.1715}, we presented global fits of the cMSSM including data from precision tests of the Standard Model (SM), constraints from cosmology on the dark matter relic abundance, LEP bounds, and constraints on SUSY from LHC 2010 data. Additionally we studied the impact of recent exclusion limit on the WIMP-proton spin-independent scattering cross-section $\sigmaSI$ from the XENON100 experiment \cite{xenon:2011hi}, a two-phase time projection chamber with a liquid xenon target, on the cMSSM parameter space. Similar studies of the impact of LHC  and XENON100 data on global fits of the cMSSM and other SUSY models are found in Refs.~\cite{arXiv:1106.2529,arXiv:1104.3572,arXiv:1105.5162,arXiv:1111.6098,arXiv:1110.3568}. 
Our approach differs from other studies, because we provide results for both a Bayesian and frequentist statistical analysis, and included for the first time the relevant uncertainties (astrophysical and related to hadronic matrix elements) for the analysis of direct detection data. 

In this work we update the global fits of the cMSSM presented in Ref.~\cite{arXiv:1107.1715} to include LHC 1 fb$^{-1}$ exclusion limits and LHC 5 fb$^{-1}$ constraints on the Higgs boson (See also Refs.~\cite{arXiv:1104.3572,arXiv:1111.6098,arXiv:1110.3568} for a discussion of the impact of LHC 2011 SUSY searches on the cMSSM, as well as on more general SUSY models. See Ref.~\cite{Buchmueller:2011ab} for a study of the impact of a hypothetical Higgs discovery at 125 GeV on the cMSSM parameter space). In our analysis, we apply all the experimental bounds used in Ref.~\cite{arXiv:1107.1715}, most notably constraints on $\sigmaSI$ derived from the XENON100 experiment. We use nested sampling as a scanning algorithm \cite{Feroz:2007kg} and study both the resulting Bayesian posterior distributions and the frequentist best fit regions. The methodology used has been described in detail in previous papers  \cite{deAustri:2006pe,Trotta:2008bp,Feroz:2011bj,arXiv:1107.1715}, in particular we closely follow the procedure presented in~\cite{arXiv:1107.1715}. We do not repeat the description of the methodology here and instead focus on the discussion of new physical results from the tighter LHC constraints. For further details on the statistical and theoretical background of our approach we refer the reader to Ref.~\cite{arXiv:1107.1715} and references therein.

Constraints on the WIMP mass and spin-independent cross-section derived from direct detection data strongly depend on the parameters describing the dark matter halo model, namely, the WIMP velocity distribution and the local density~\cite{ST}. Additionally, constraints derived on the cMSSM parameters from the XENON100 exclusion limit depend on the hadronic matrix elements, which parameterise the contributions of light quarks to the proton composition. None of these quantities is precisely known, and neglecting the uncertainty on these quantities in the modelling can lead to incorrect inferences for the cMSSM parameters. Our inference procedure consistently takes into account our (lack of) knowledge of these parameters. We account for both astrophysical and nuclear physics uncertainties entering in the calculation of the event rate in direct detection experiments, by treating them as nuisance parameters in the analysis, which in the end are marginalised/maximised over to correctly infer the cMSSM parameters. 

This paper is organised as follows. In Section 2 we present the theoretical and statistical framework for the analysis. In Section 3 we present the results, Section 4 is devoted to the discussion of the prospects for discovering SUSY in the cMSSM in different detection channels. We present our conclusions in Section 5.

\section{Theoretical and statistical framework}
\label{sec:theory}

The cMSSM is described by four continuous parameters:
the universal mass terms for scalars $m_0$ and gauginos $m_{1/2}$, the universal scalar trilinear coupling $A_0$ and the ratio of the Higgs vacuum expectation values $\tan \beta$. 
There is an additional discrete parameter, the sign of the Higgsino mass parameter, which is fixed to sgn$(\mu) = +1$ in the following, since this is the value favoured by current measurements of the anomalous magnetic moment of the muon~\cite{Feroz:2008wr}.
Despite its small number of parameters the cMSSM contains several of the main phenomenological features of SUSY. Additionally the Lightest Supersymmetric Particle (LSP) is usually the lightest neutralino $\neut$, a superposition of the neutral gauginos and higgsinos, which is an excellent cold dark matter candidate. This is a consequence of the structure of the Renormalization Group Equations, which tend to make the Bino mass parameter small at the electroweak scale. 
Similarly, the stau is the lighter of all the sleptons, and in fact, it can become the LSP if the universal scalar mass is small with respect to the gaugino mass. In the regions where the stau is only slightly heavier than the neutralino a coannihilation effect in the early Universe helps reducing the neutralino relic density, making it easier to reproduce the observable dark matter abundance.
Finally, for large values of the trilinear parameter the lighter stop can also become the LSP or provide a similar coannihilation effect to the relic density of neutralinos \cite{Ellis:2001nx}.

Residual uncertainties on the values of some SM parameters can strongly influence the constraints derived on the model parameters and observables~\cite{Roszkowski:2007fd}. Therefore, four SM nuisance parameters (namely, the top and bottom masses and the electroweak and strong coupling constants) are included in the scan. Additionally, when including constraints from direct detection searches in the analysis, large uncertainties enter via astrophysical and nuclear physics quantities. In Ref.~\cite{arXiv:1107.1715} we provided a detailed discussion on the origin of these uncertainties and their impact on the calculation of the direct detection recoil rate. In the following we simply recall the resulting parameterisation of the uncertainties and refer the reader to the above paper for further details. 

The main astrophysical uncertainties are quantified using four nuisance parameters, describing both the abundance of dark matter and the WIMP velocity distribution. We use a simple Ansatz to describe the velocity distribution function (see section 3.2 in Ref.~\cite{arXiv:1107.1715}). This parameterisation can correctly reproduce the different WIMP halo profiles that result from N-body simulations \cite{Lisanti:2010qx}. However, it is important to keep in mind that the actual local dark matter velocity distribution function remains unknown. N-body simulations have shown that the velocity distribution strongly depends on the detailed assembly history of our galaxy \cite{Vogelsberger:2008qb}. Additionally, for simplicity we neglect the presence of a dark disk \cite{Read:2008fh,Bruch:2008rx,Purcell:2009yp} in our modelling of uncertainties. We do not include microhalos and localised velocity streams in our analysis, since the impact of these on direct detection signals has been estimated to be small (see e.g. Ref.~\cite{Vogelsberger:2008qb,Schneider:2010jr}). For a discussion of the different sources of astrophysical uncertainties affecting direct detection experiments see e.g. Ref.~\cite{Serpico:2010ae}.

The hadronic uncertainties arise  when translating constraints on the cMSSM parameters into constraints on the WIMP-proton spin-independent cross-section $\sigmaSI$. This cross-section depends on the hadronic matrix elements $f_{T_q} = m_p <p|m_q \bar{q} q|p>$, which represent the contributions of the light quarks $q = \{u,d,s\}$ to the proton composition. These quantities are not precisely known and are treated as nuisance parameters in the analysis. Therefore, in addition to the four cMSSM model parameters, we also scan over up to eleven nuisance parameters, thereby fully accounting for those astrophysical and hadronic uncertainties. The specific ranges used for all the above nuisance parameters can be found in Table~II of Ref.\,\cite{arXiv:1107.1715}.

We investigate the constraints that can be placed on the cMSSM parameters using Bayesian methods to explore both the posterior probability density function (pdf) and the profile likelihood~\cite{Feroz:2011bj}. Bayesian inference is based on Bayes' theorem~\cite{Trotta:2008qt}
\begin{equation}
p(\params|\mathbf{D})=\frac{p(\mathbf{D}|\params) p(\params)}{p(\mathbf{D})},
\label{eqn:Bayes}
\end{equation}
which states that, given a likelihood function $p(\mathbf{D}|\params)$ for a set of experimental data $\mathbf{D}$ and parameters of interest $\params$, and some prior pdf $p(\params)$ representing any information about the parameters available before taking into account the data, one can find the posterior probability $p(\params|\mathbf{D})$ for the parameters. For the purpose of parameter inference the Bayesian evidence $p(\mathbf{D})$ acts as a normalisation constant, and is not important in the following analysis. As can be seen from Eq.~\eqref{eqn:Bayes} the posterior distribution depends on both the likelihood function and the prior. In the ideal case the experimental data are constraining enough to overcome the effect of the prior, such that the posterior distribution will be dominated by the likelihood function. When applying Bayes' theorem to complicated multi-dimensional models the data is often not sufficiently constraining to completely overcome the priors. By repeating the analysis for different choices of prior distributions one can assess to which extent the posterior pdf is dominated by the experimental constraints. Should the posterior display some residual prior dependence one has to be careful with the interpretation of the resulting constraints on the observables~\cite{Trotta:2008bp,Scott:2009jn}. Such a prior dependence has for example been observed in \cite{arXiv:1107.1715}, where two different sets of priors were applied. We thus repeat each of our scans for two different sets of priors: ``flat'' priors (uniform in the cMSSM masses), and ``log" priors (uniform in the log of the cMSSM masses). Both sets of priors are uniform on $A_0$ and $\tan \beta$. For the nuisance parameters we choose informative Gaussian priors, with the mean and standard deviation chosen appropriately to reflect current experimental constraints. The range covered by the scan for the cMSSM parameters, and the mean and standard deviation for the Gaussian priors on the nuisance parameters are given in Table~I and Table~II of Ref.~\cite{arXiv:1107.1715}, respectively. For scans that do not include astrophysical and hadronic uncertainties the nuisance parameters were fixed to their mean value. 

In order to identify constraints derived on a subset of one or two parameters one can consider two different quantities. The Bayesian marginalised one-dimensional posterior pdf for some parameter $\Theta_i$ can be found from the full posterior pdf by integrating out the remaining parameters
\begin{equation}
p(\Theta_i|\mathbf{D})=\int p(\params|\mathbf{D}) d\Theta_1 ... d\Theta_{i-1} d\Theta_{i+1} \Theta_{n},
\end{equation}
and equivalently for the 2-dimensional marginalised posterior pdf. Alternatively one can consider the frequentist profile likelihood function. The one-dimensional profile likelihood function for some parameter $\Theta_i$ can be found by maximising over all other parameters
\begin{equation}
{\mathcal L}(\Theta_i)= \max_{\Theta_1,...,\Theta_{i-1},\Theta_{i+1},...,\Theta_{n}}\mathcal{L}(\params).
\end{equation}
These two statistical quantities have a different meaning and may lead to different conclusions. The profile likelihood function is ideal to discover small regions in parameter space that correspond to a large likelihood value. In contrast, the marginalised posterior pdf accounts for volume effects by integrating over hidden dimensions, and thus peaks around the region of highest posterior mass. Both of these quantities provide valuable information about the parameter space of interest, therefore we present results for both the marginalised posterior pdf and the profile likelihood function. To obtain the posterior distribution we modified the publicly available \texttt{SuperBayeS v1.5.1} package \cite{SuperBayeS}\footnote{For this paper, the public \texttt{SuperBayeS v1.5.1} code has been modified to interface with with DarkSUSY 5.0.5~\cite{DarkSUSY,Gondolo:2004sc}, SoftSUSY 2.0.18 
\cite{SoftSUSY,Allanach:2001kg}, MicrOMEGAs 2.0  \cite{MicrOMEGAs,Belanger:2006is}, SuperIso 2.4 \cite{SuperIso,Mahmoudi:2008tp} and SusyBSG 1.4 \cite{SusyBSG,Degrassi:2007kj}.}.
As a scanning algorithm we use MultiNest v2.8~\cite{Feroz:2007kg,Feroz:2008xx}, the running parameters are the same as in Ref.~\cite{arXiv:1107.1715}. Each scan is repeated for both log and flat priors. The profile likelihood, which is in principle prior independent, is derived from combined chains of these two scans, as advocated in Ref.~\cite{Feroz:2011bj}.The chains used to derive the constraints on the cMSSM parameters and the observables are generated from approximately 7 (43) million likelihood evaluations, with an overall efficiency of 0.1 (0.02) for fixed (varying) astrophysical and nuclear physics nuisance parameters.

The experimental data included in the likelihood function are given in Table~III of Ref. \cite{arXiv:1107.1715}. The full likelihood function includes precision tests of the SM, constraints on the dark matter relic density from 7-year WMAP data and constraints from accelerator searches. We also include recent data from the XENON100 direct detection experiment, obtained with an effective volume of $48$ kg and an exposure of $100.9$ days between January and June 2010 ~\cite{xenon:2011hi}. Three candidate scattering events were detected within the signal region, which is compatible with the expected number of background events $b = 1.8 \pm 0.6$, resulting in tight exclusion limits in the $(m_{\neut},\sigmaSI)$ plane. Our full likelihood function is described in detail in Ref.  \cite{arXiv:1107.1715}, which we refer the reader to for further details.

In this paper we upgrade the constraints from SUSY searches at the LHC to include recent exclusion limits from 1 fb$^{-1}$ integrated luminosity data, as well as the latest limits on the SM Higgs mass obtained with 5 fb$^{-1}$ integrated luminosity worth of data. We focus on constraints presented in Ref.~\cite{arXiv:1111.4820} by the CMS collaboration, which currently provides the most stringent constraints on SUSY. The CMS data set was collected for a centre-of-mass energy of $\sqrt{s}=7$ GeV and an integrated luminosity of 1 fb$^{-1}$. A SUSY signal is searched for in hadronic events with two or more jets and missing transverse energy using the kinematic variable $\alpha_T$. No significant signal beyond the SM predictions was observed, and so lower limits on the cMSSM masses $m_0$, $m_{1/2}$ were derived, as shown in Fig.~5 of Ref.~\cite{arXiv:1111.4820}.  While this exclusion curve was obtained for fixed values $\tan \beta = 10$, $A_0 = 0$, the decay channels used to obtain this limit are relatively insensitive to the values of these two parameters. Thus, we can treat the LHC constraint as approximately independent of $\tan \beta$ and $A_0$~\cite{Allanach:2011ut}. The LHC 2011 exclusion limit is included in the likelihood function by defining the likelihood of samples corresponding to masses below the limit to be zero.

The ATLAS and CMS collaborations have recently announced new stringent constraints on the Standard Model Higgs boson mass, $115.5 < m_h < 131$ GeV (at 95\%, from a combined 11 channels analysis of ATLAS data), derived from searches with 5 fb$^{-1}$ integrated luminosity. Furthermore, a $\sim 3.6\sigma$ local excess has been reported for $m_h \sim 126$ GeV; the significance is reduced to $\sim 2.3\sigma$ when the ``look-elsewhere'' effect is considered. Although it is too early to interpret such an excess as an indication of the Higgs, we include the updated lower limit in our analysis, by applying the 95\% lower limit of $m_h > 115.5$, with an added theoretical uncertainty of 3 GeV, according to the procedure described in Ref.~\cite{deAustri:2006pe}. Although the ATLAS limit applies to the Standard Model Higgs, the lightest Higgs in the cMSSM is almost invariably SM-like, hence we can adopt the above limit in this context. 


\section{Results}
\label{secresults}

The results of our numerical analysis will be presented in three steps. First we show the impact on the cMSSM parameter space resulting from all the constraints described in the previous section (except those from direct detection experiments), including the latest LHC data. Afterwards we combine these results with XENON100 data, fully marginalising/maximising over the astrophysical and hadronic nuisance parameters. Finally, we discuss in some detail the impact of the experimental constraint on the anomalous magnetic moment of the muon on the cMSSM parameter space.

\begin{figure*}
\centering
\includegraphics[width=0.32\linewidth]{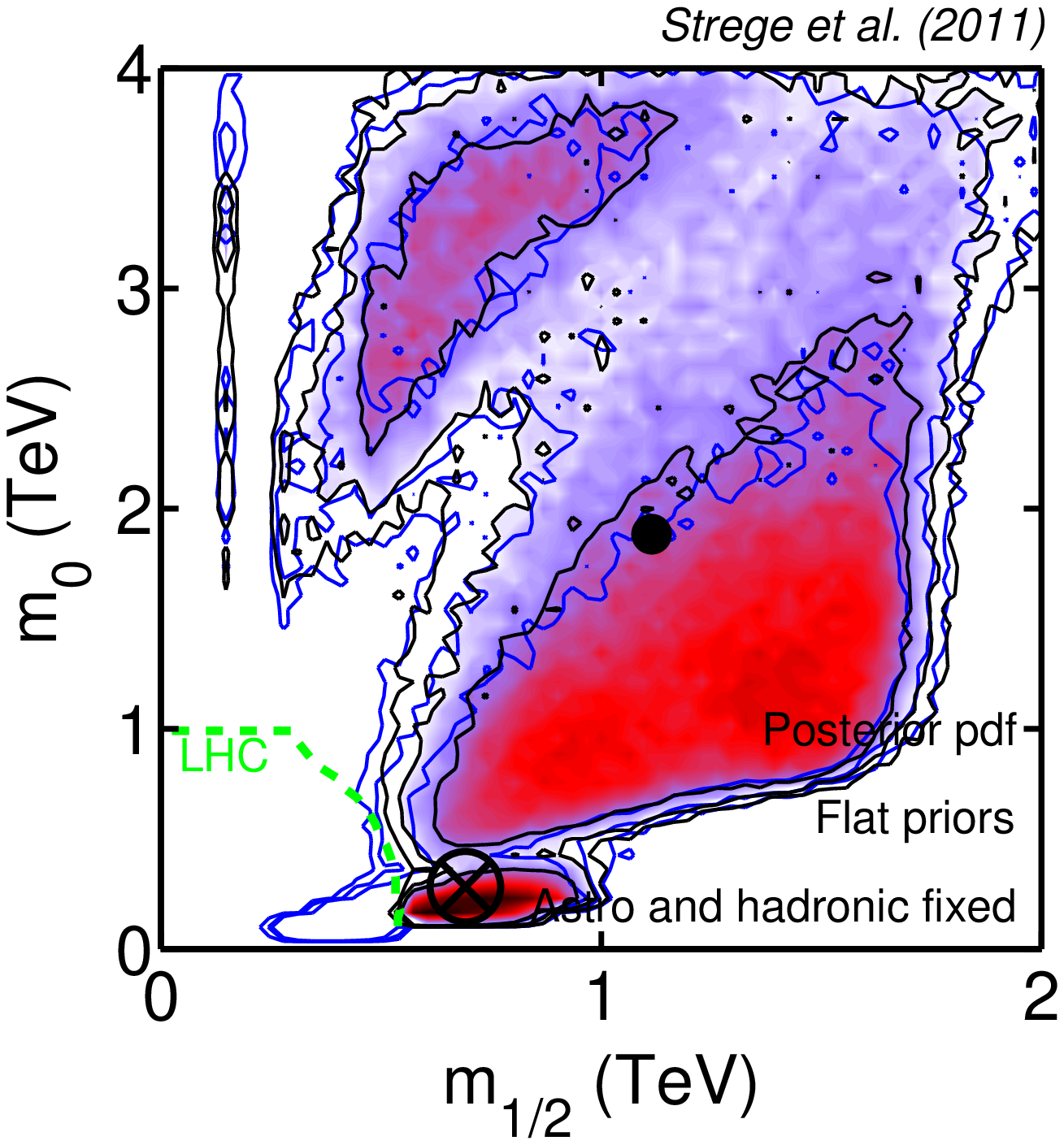}
\includegraphics[width=0.32\linewidth]{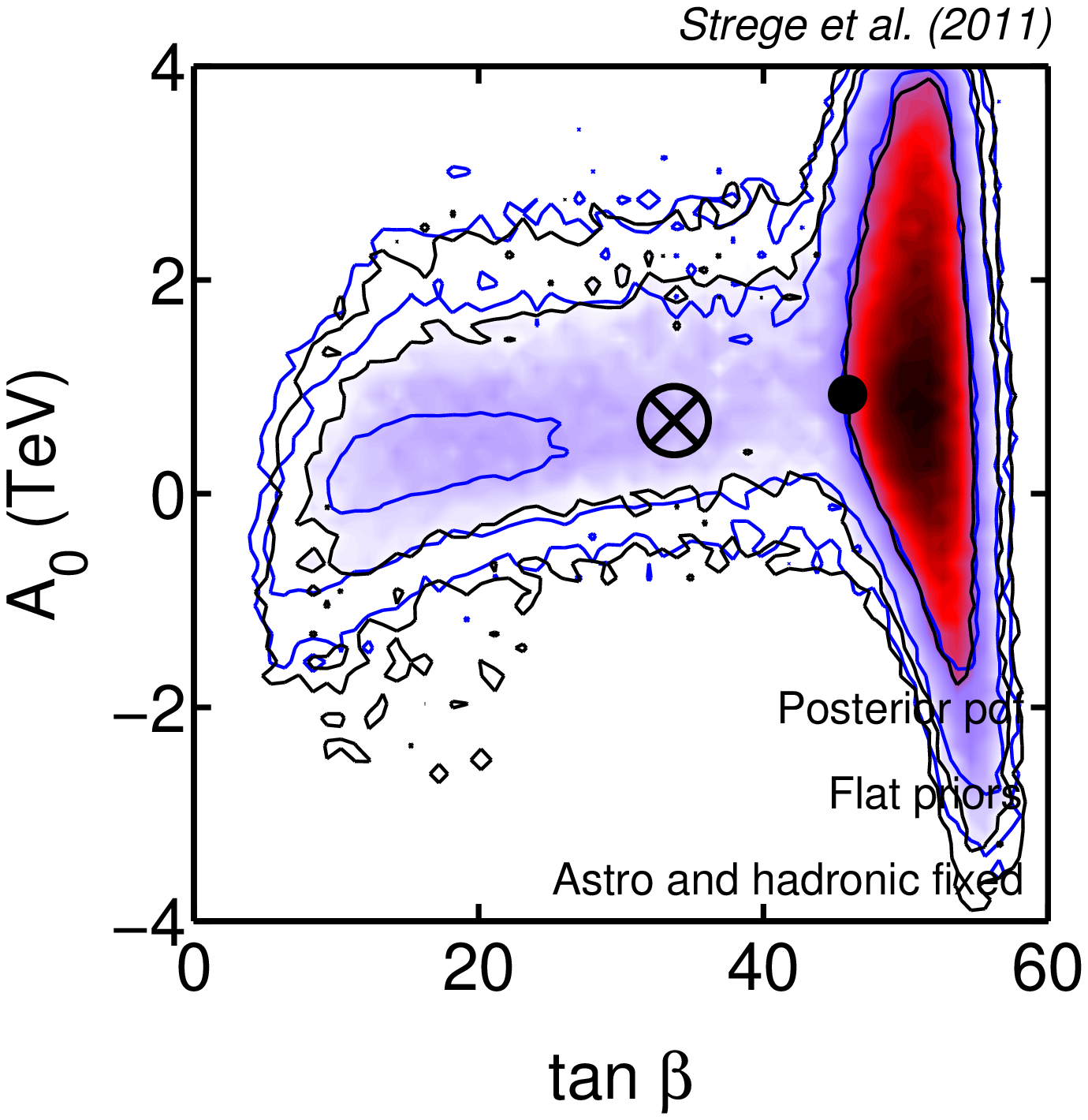}
\includegraphics[width=0.32\linewidth]{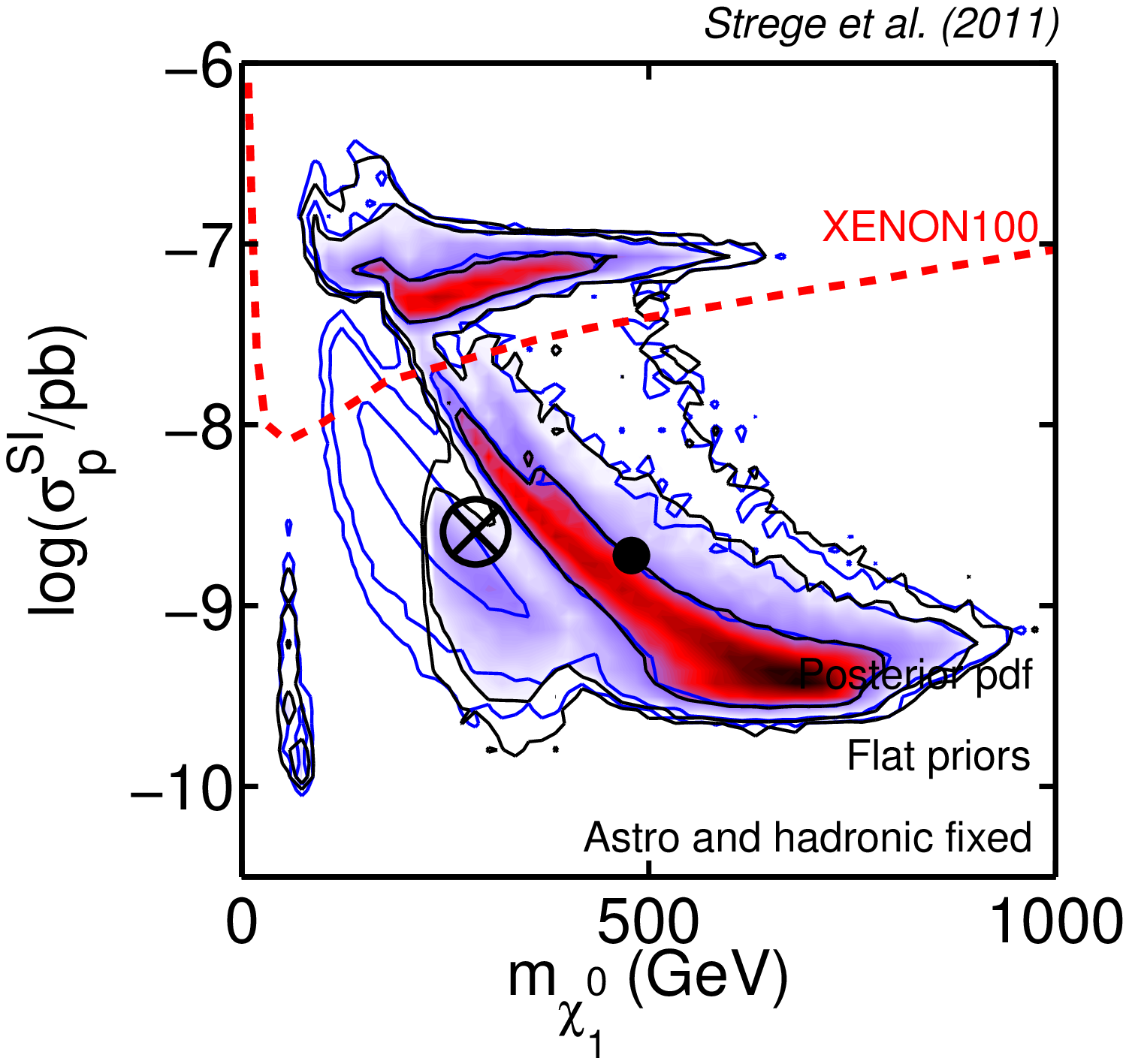} \\
\includegraphics[width=0.32\linewidth]{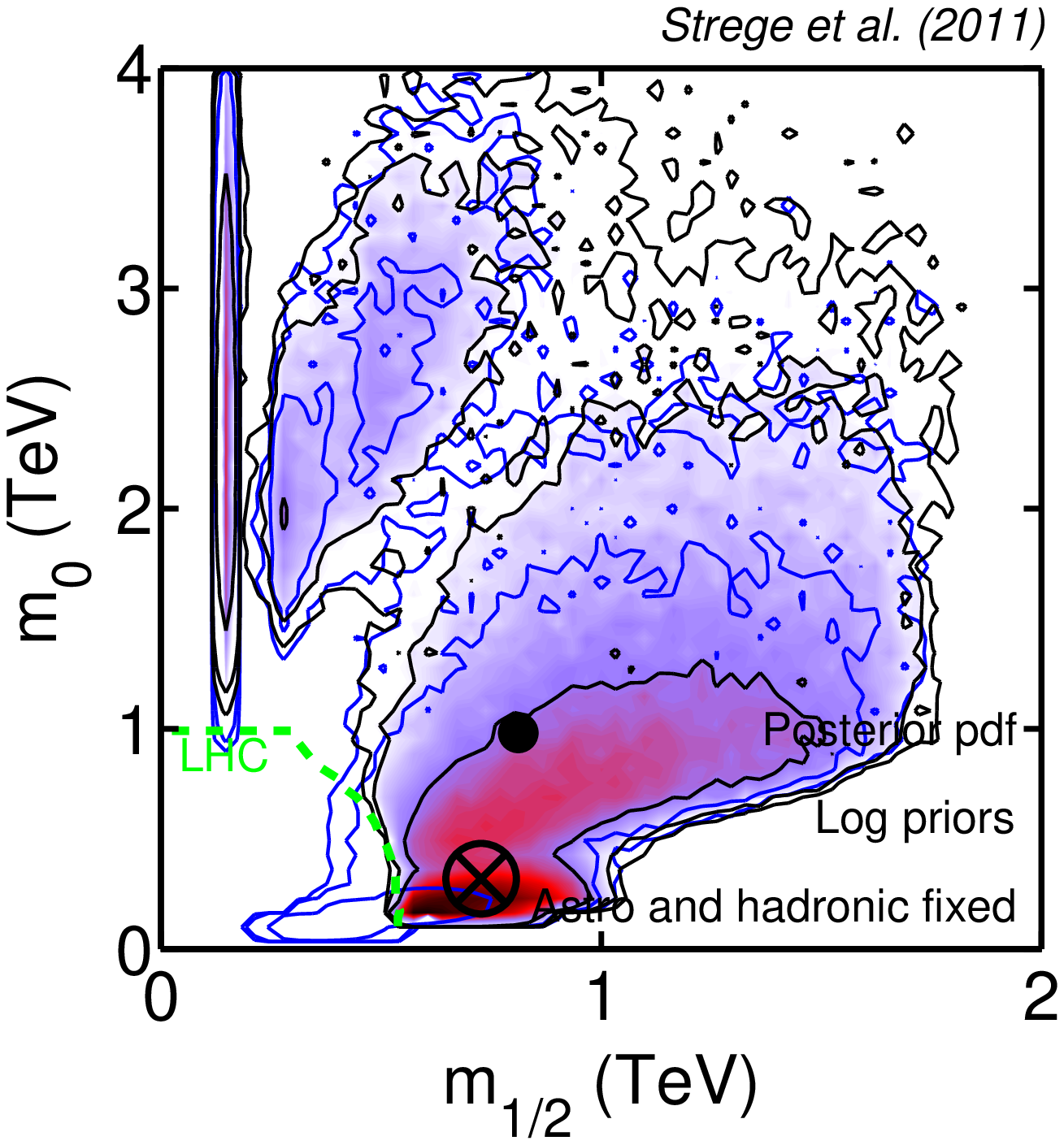}
\includegraphics[width=0.32\linewidth]{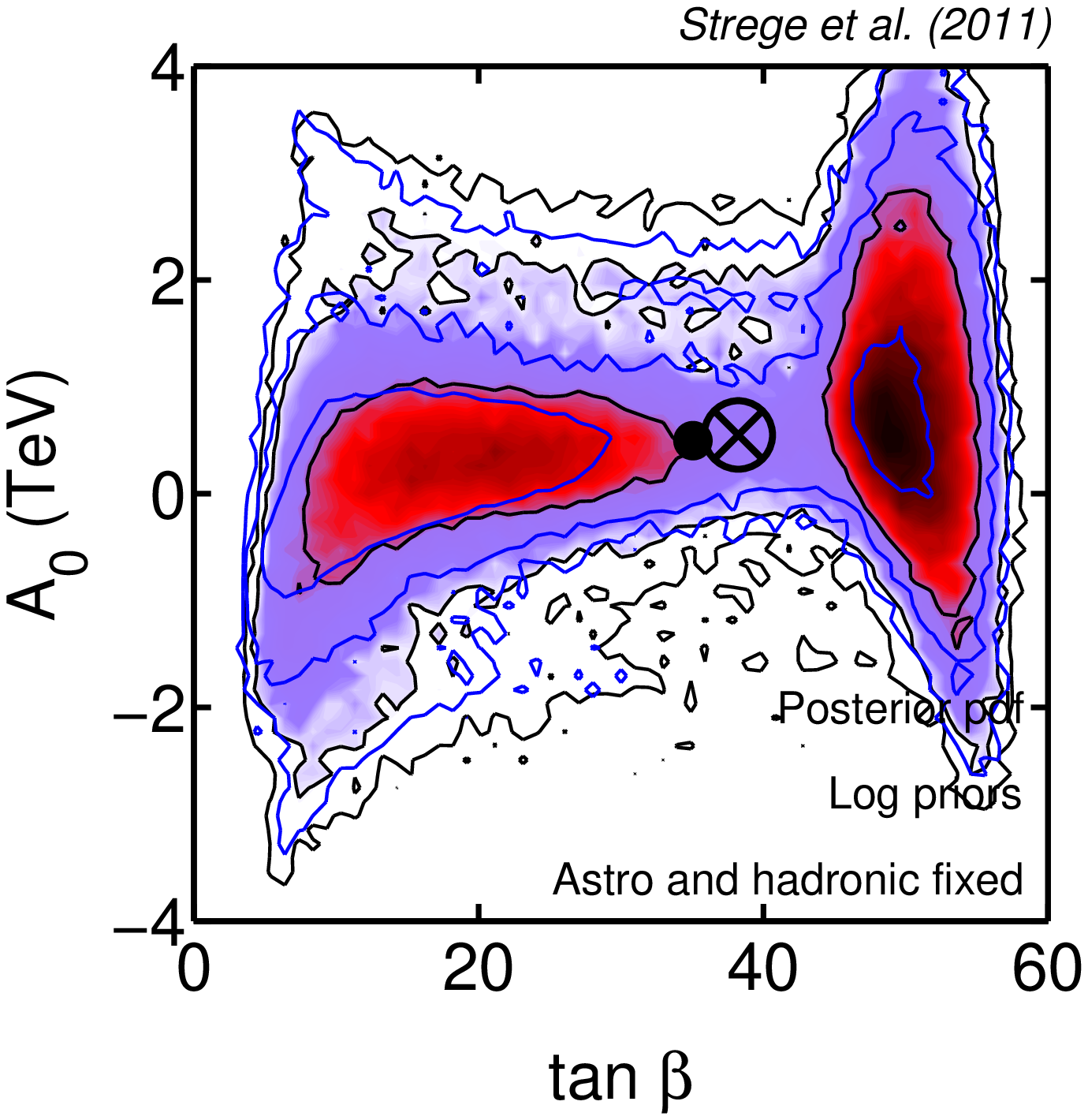}
\includegraphics[width=0.32\linewidth]{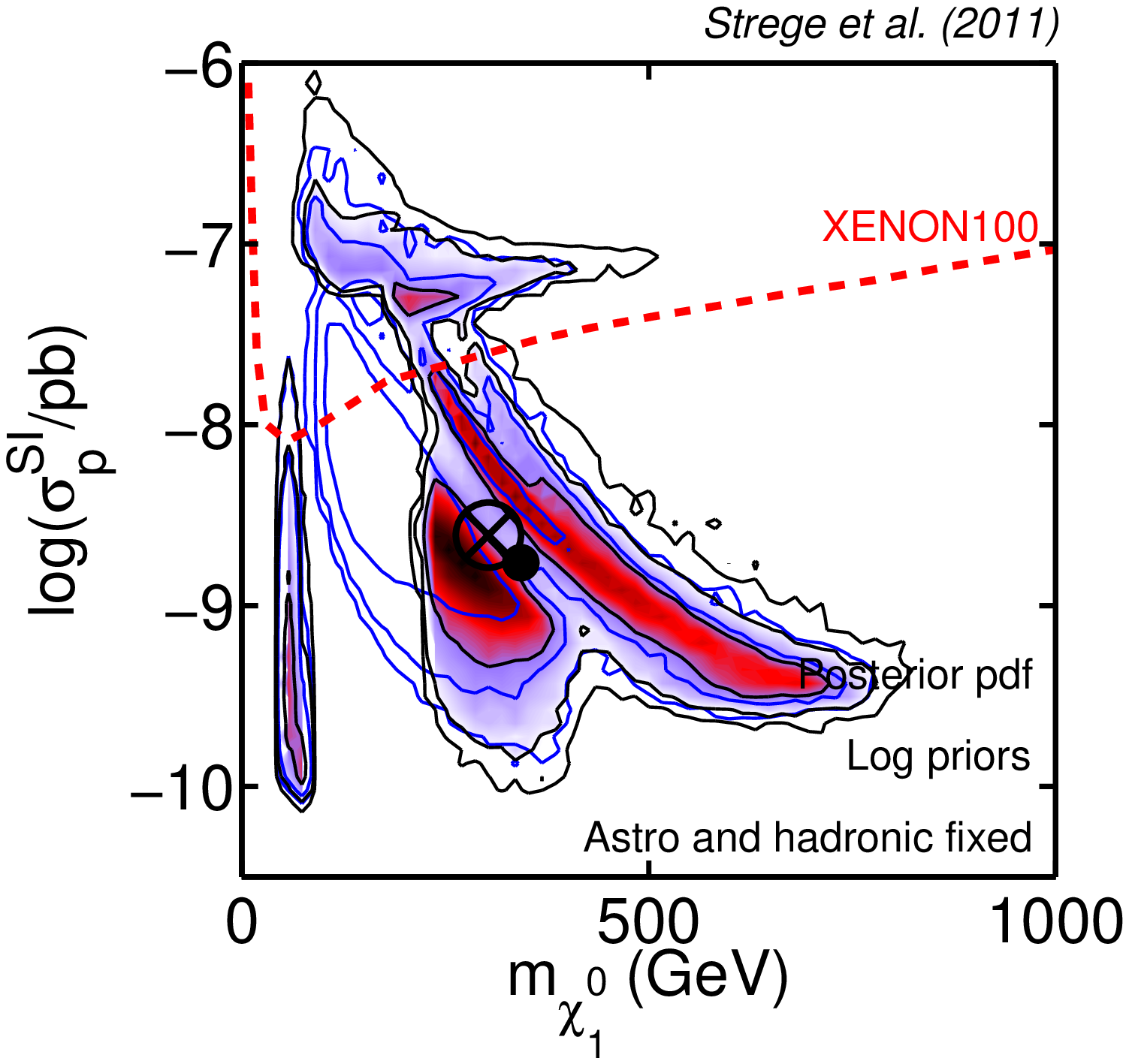}
\includegraphics[width=0.32\linewidth]{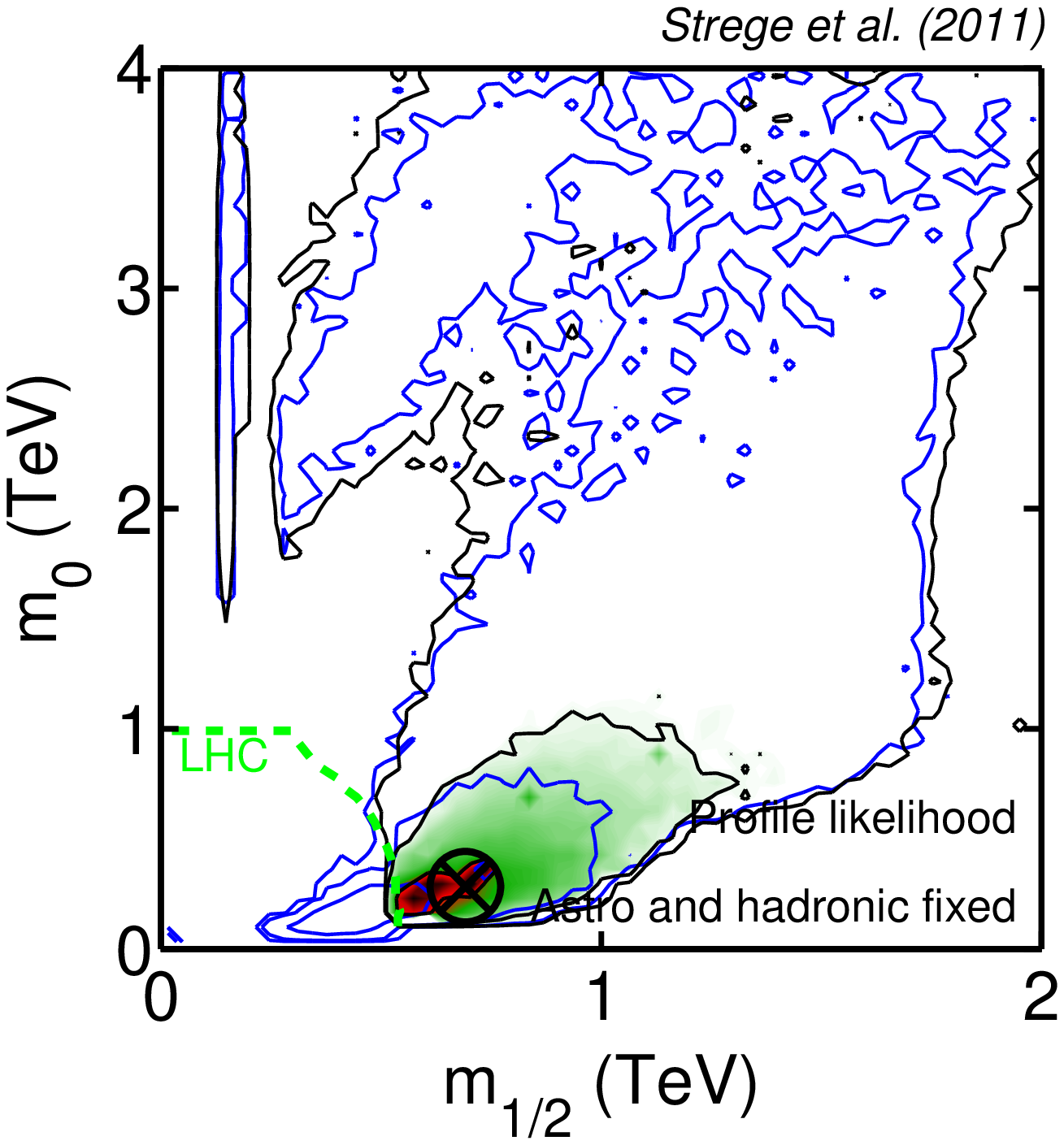}
\includegraphics[width=0.32\linewidth]{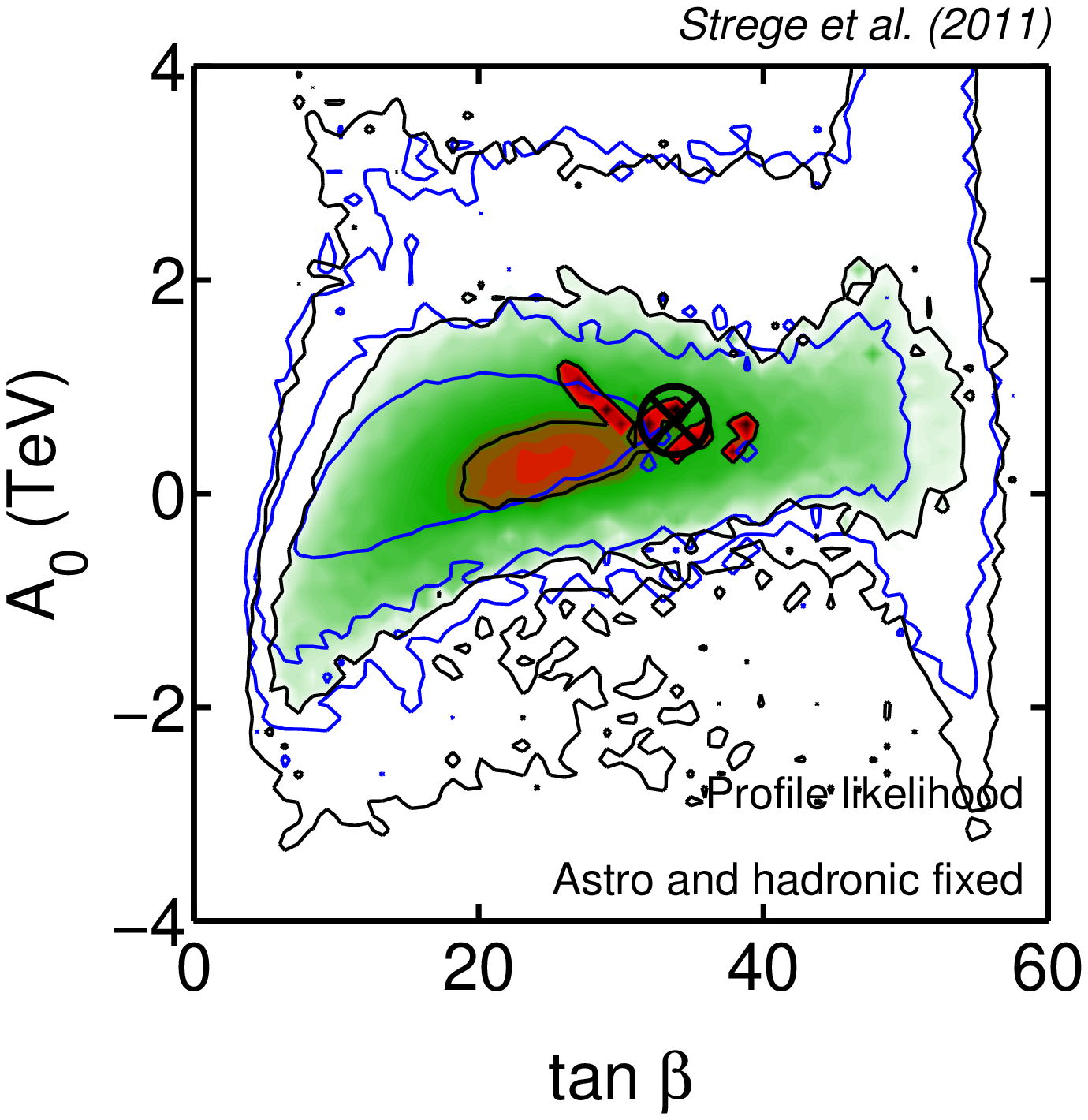}
\includegraphics[width=0.32\linewidth]{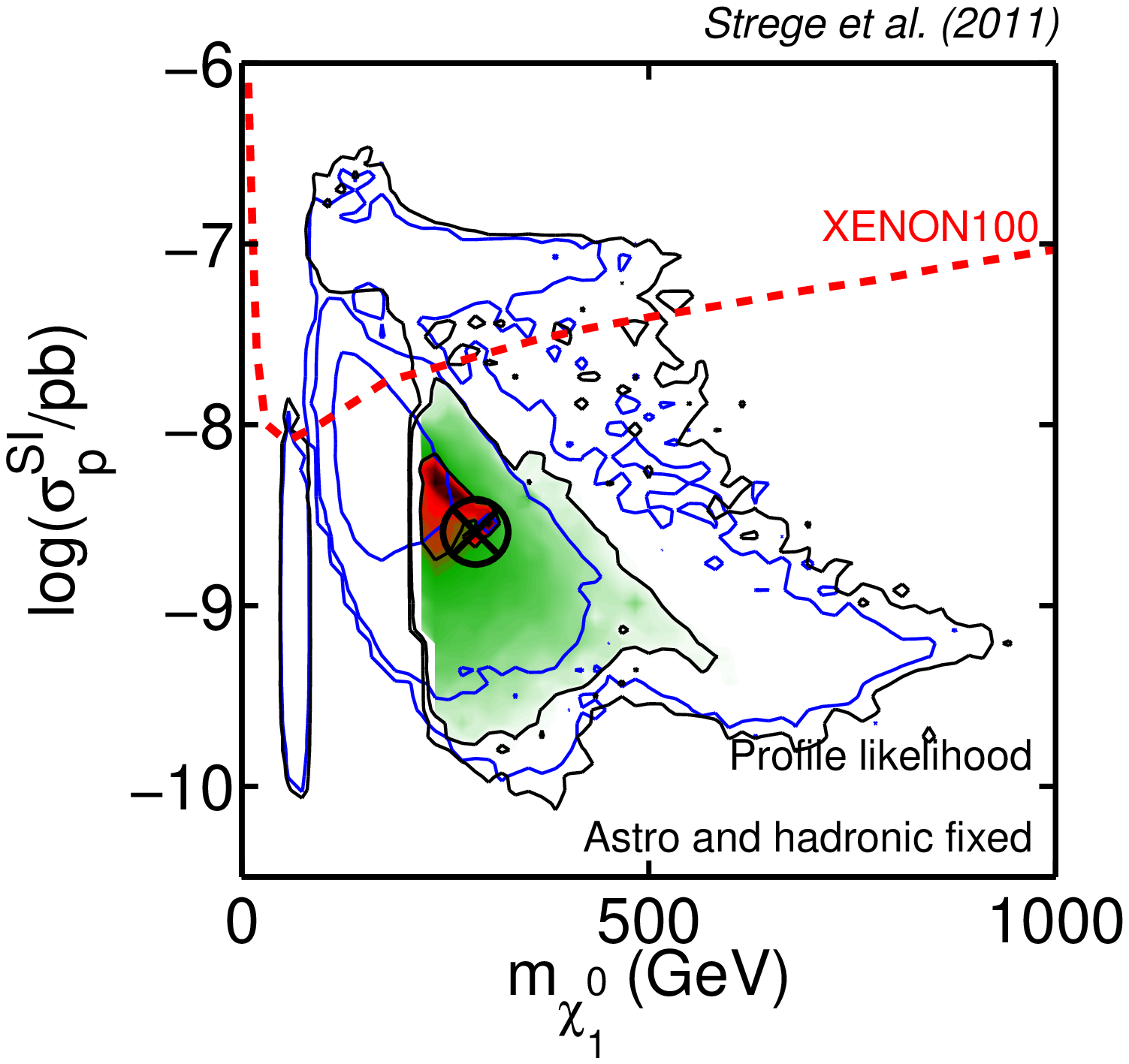}
\caption{\fontsize{9}{9} \selectfont Black contours: posterior pdf (upper panels, for flat and log priors) and profile likelihood (lower panels) for the cMSSM parameters, including all present-day constraints (WMAP 7-years, LHC 1 fb$^{-1}$ SUSY searches and 5 fb$^{-1}$ Higgs limits included), except XENON100. From the inside out, contours enclose 68\%, 95\% and 99\% of  marginal posterior probability (top two rows) and the corresponding profiled confidence intervals (bottom panels). The black cross represents the best fit point, the black dot the posterior mean (for the pdf plots). Parameters describing astrophysical and hadronic uncertainties have been fixed to their fiducial values. Blue contours represent the constraints obtained without the inclusion of LHC data. In the plots on the left, the dashed/green line represents the current LHC exclusion limit, while in the right-most plots the red/dashed line is the 90\% exclusion limit from XENON100, from Ref.~\cite{xenon:2011hi}, rescaled to our fiducial local DM density of $\rho_\text{loc} = 0.4$ GeV/cm$^3$. \label{fig:ALL_NoXe_LHC}}  
\end{figure*}

\begin{table*}
\begin{center}
\begin{tabular}{| l | c | c | c |}
\hline
& LHC 2011 &  LHC 2011 + XENON100 & LHC 2011 w/o $\delta a_\mu^{SUSY}$ \\
\hline
\multicolumn{4}{|c|}{Input parameters}\\\hline
$ m_0$ [GeV] & 282.19 & 267.54 & 188.80 \\
$ m_{1/2}$ [GeV] & 691.76 & 635.56 & 908.06 \\
$A_0$ [GeV] & 685.35 & 935.04 & -630.62 \\
$ \tan\beta$ & 33.74 & 29.75 & 8.54 \\
\hline 
$M_t$ [GeV] & 174.108 & 173.397 & 173.338 \\
$m_b(m_b)^{\bar{MS}}$ [GeV] & 4.214 & 4.221 & 4.217 \\
$[\alpha_{em}(M_Z)^{\bar{MS}}]^{-1}$ & 127.952 & 127.952 & 127.955 \\
$\alpha_s(M_Z)^{\bar{MS}}$ & 0.116 & 0.119 & 0.117 \\ \hline
\multicolumn{4}{|c|}{Observables}\\\hline
$m_h$ [GeV] & 117.1 & 115.6 & 119.4 \\
$m_\neut$ [GeV] & 287.2  & 262.2 & 382.5 \\
$\delta a_\mu^{SUSY} \times 10^{10}$ & 22.82 &  32.57 & 2.32 \\ 
$BR(\bar{B} \rightarrow X_s\gamma) \times 10^4$ & 3.04 & 3.42 & 3.36 \\
$\DeltaO  \times 10^{2}$ & 8.42 & 8.36 & 8.04 \\ 
$\Dstaunu \times 10^{2}$ & 4.81 & 4.81 & 4.82 \\
$\Dsmunu \times 10^{3}$ & 4.97 & 4.97 & 4.98 \\
$\Dmunu \times 10^{4}$ & 3.86 & 3.86 & 3.86 \\
$\sigmaSI$ [pb] &  $2.6 \times 10^{-9}$ & $3.8 \times 10^{-9}$ & $3.0 \times 10^{-10}$\\
$\sigmaSD$ [pb] & $3.8 \times 10^{-7}$  & $5.8 \times 10^{-7}$ & $3.7 \times 10^{-8}$\\
$\Omega_\neut h^2$ & 0.1174 & 0.1121 & 0.1118 \\ 
\hline
\end{tabular}
\end{center}
\caption{Input cMSSM parameters (and some other interesting quantities) for the best fit points. The column ``LHC 2011'' is for the case where LHC 2011 data, but no direct detection data is applied, while ``LHC + XENON100'' denotes the case where XENON100 data is added to the analysis and astrophysical and hadronic nuisance parameters have been included in the scan and profiled over. The column ``LHC 2011 w/o $\delta a_\mu^{SUSY}$" is for the case where all constraints, including LHC 2011 data, except for direct detection data and the $\delta a_\mu^{SUSY}$ constraint, are applied. \label{tab:bestfit}}
\end{table*}

\begin{table*}
\begin{center}
\begin{tabular}{| l | c | c | c |}
\hline
& LHC 2011 & LHC 2011 + XENON100 & LHC 2011 w/o $\delta a_\mu^{SUSY}$ \\\hline 
Observable & & & \\
\hline

SM nuisance & 1.059 & 0.528 & 0.100 \\
Astro nuisance &  N/A & 1.052 & N/A \\
Hadronic nuisance & N/A & 0.172 & N/A \\
$M_W$ & 0.751 & 1.074 & 1.228 \\
$\sin^2\theta_{eff}$ & 0.158 & 0.084 & 0.044 \\
$\delta a_\mu^{SUSY} $ & 0.660 & 0.127 & 0.000 \\
$BR(\bar{B} \rightarrow X_s\gamma) $ & 1.665 & 0.113 & 0.218 \\
$\Delta M_{B_s}$ & 0.331 &  0.243 & 0.306 \\
$\RBtaunu$  &  0.918 & 0.867 & 0.551 \\
$\DeltaO$  & 3.305 & 3.232 & 2.812 \\
$\RBDtaunuBDenu $ & 0.868 & 0.859 & 0.799 \\
$\Rl$ & 0.390 & 0.382 & 0.328 \\
$\Dstaunu $ & 2.252 & 2.248 & 2.214 \\
$\Dsmunu  $ & 3.136 & 3.131 & 3.098 \\
$\Dmunu $  & 0.008  & 0.008 & 0.008 \\
$\Omega_\neut h^2$ & 0.171 & 0.0002 & 0.002 \\ \hline
$m_h$ & 1.07 & 2.157 & 0.287 \\
XENON100 & N/A & 0.019 & N/A \\
LHC & 0.0 & 0.0 & 0.0 \\ 
Sparticles (LEP) & 0.0 & 0.0 & 0.0 \\ 
$\brbsmumu$ & 0.0 & 0.0 & 0.0 \\ \hline
Total & 16.74 & 16.29 & 12.00 \\
$\chi^2/$dof & 1.86 & 1.81  & 1.50 \\
\hline
\end{tabular}
\end{center}
\caption{Breakdown of the total $\chi^2$ by observable for the best fit points, for both the case where all data, including LHC 2011 constraints, except for XENON100 data were applied (left column), when all data including XENON100 results were applied and nuisance parameters were profiled over (central column), and when all data including LHC 2011 constraints except for XENON100 data and the $\delta a_\mu^{SUSY}$ constraint were applied (right column). \label{tab:chisquare}}
\end{table*}

\subsection{Impact of LHC 2011 data}
\label{sec:LHC_only}

In Fig.~\ref{fig:ALL_NoXe_LHC} we show the impact of LHC 2011 data on the cMSSM in the $(m_{1/2},m_0)$ plane (left), the $(\tan \beta,A_0)$ plane (centre) and the $(m_{\neut},\sigmaSI)$ plane (right). The two-dimensional marginalised posterior pdfs are given for both flat (top panel) and log (central panel) priors. The profile likelihood is shown in the bottom panel. Blue contours correspond to constraints derived without including LHC data, black contours show constraints on the parameters after the latest LHC data are included. 

For the pre-LHC contours we observe several regions in the $(m_0, m_{1/2})$ plane that are of particular interest. The first is the so-called focus point (FP) region \cite{Feng:1999zg,Feng:2000gh}, or hyperbolic branch \cite{Chan:1997bi}, which corresponds to a region in parameter space where the Higgsino mass parameter $\mu$ is very small, while scalar masses can be very large. Due to the smallness of $\mu$ the neutralino has a large higgsino component, which leads to its relic density to decrease \cite{Baer:1997ai}, such that current cosmological constraints on the dark matter relic abundance can be satisfied. In the $(m_{1/2},m_0)$ plane it corresponds to a large area at sizable $m_0 > 1$ TeV and relatively small $m_{1/2}$. 

A second region of interest is the stau co-annihilation (SC) region. In this region the lightest stau is slightly heavier than the neutralino LSP. As a result, the neutralino relic density is reduced by neutralino-stau co-annihilations, so that the WMAP constraint can be satisfied. 
For fixed $A_0$ and $\tan \beta$ the SC region appears as a narrow band on the $(m_{1/2},m_0)$ plane spanning a range of gaugino masses $m_{1/2} \sim$ a few hundred GeV at small scalar masses. For varying $A_0$ and $\tan \beta$ this region appears slightly smeared.

A third possibility for neutralinos to reproduce the correct relic abundance is for them to undergo a resonant annihilation mediated by a
relatively light pseudoscalar Higgs. This is possible when $m_{\neut}\approx 2 m_{A^0}$ and typically happens for large values of $\tan\beta$ and relatively large values of both the gaugino and scalar mass parameters. The position of this so-called Higgs funnel region in
the  $(m_{1/2},m_0)$ plane strongly depends on other parameters (in particular on $\tan\beta$ and the trilinear term).

Finally, there is a narrow viable region in which the lightest neutralino mass is approximately half of the lightest Higgs mass, $m_{\neut}\approx60$~GeV. In this case, the $s$-channel Higgs exchange in the neutralino annihilation cross section in the early Universe is resonant and the resulting relic abundance is significantly reduced.
This appears in Fig.~\ref{fig:ALL_NoXe_LHC} as a narrow area with $m_{1/2}\approx100$~GeV and spanning several orders of magnitide in $m_0$. 
For a further discussion of the features of the cMSSM see e.g. Ref.~\cite{Trotta:2008bp}.

The LHC 2011 data has a strong impact on the posterior distribution in the $(m_{1/2},m_0)$ plane, as shown on the left-hand side of the upper and central panels in Fig.~\ref{fig:ALL_NoXe_LHC}. The LHC 2011 exclusion limit cuts deep into the stau co-annihilation region. A large region that was previosuly favoured at the $68\%$ and $95\%$ level (see Ref.~\cite{arXiv:1107.1715}) is disfavoured with high confidence and the posterior contours are pushed towards higher values of $m_0$ and $m_{1/2}$. Notice that the narrow band with resonant Higgs annihilation remains viable for neutralinos with a mass around 60\,GeV.

As can be seen on the right-hand side of Fig.~\ref{fig:ALL_NoXe_LHC} the LHC exclusion limit also has important consequences for direct detection of the cMSSM: The exclusion of small gaugino masses leads to small $m_{\neut}$ being strongly disfavoured, such that a large part of previously favoured cMSSM parameter space at small and intermediate WIMP masses is now excluded at the 99$\%$ level.  In the $(m_{\neut},\sigmaSI)$ plane the FP region appears as a large island corresponding to relatively large $\sigmaSI$, lying above the projected XENON100 90$\%$ exclusion limit shown in red (not applied here). One can already see from this limit that adding direct detection data has the capability of ruling out significant portions of the FP region, which appears at high scalar masses and is therefore unaffected by current LHC constraints.

In the central plot one can see that LHC 2011 constraints have a minimal impact on the credible regions for $\tan\beta$ and $A_0$. This is to be expected because, as stated above, the LHC exclusion limit is fairly independent of the precise values of these two parameters.

These results can be compared to the impact of earlier LHC 35 pb$^{-1}$ limits on global fits of the cMSSM, as presented in Ref.~\cite{arXiv:1107.1715}. Qualitatively the resulting posterior distributions are quite similar. The main difference is that while the 35 pb$^{-1}$ exclusion limit had a fairly modest impact on the cMSSM parameter space, the LHC 2011 data leads to exclusion of higher values of the scalar and gaugino masses, so that a large portion of the stau-coannihilation region is ruled out. LHC 2011 data also impacts on the posterior distribution in the $(m_{\neut},\sigmaSI)$ plane, which was more or less unaffected by the 35 pb$^{-1}$ data. Now a significant range of small WIMP masses is excluded at low and intermediate $\sigmaSI$.

The above observations are true for both the log and the flat prior case. Due to ``volume effects" the flat prior contours extend to larger scalar and gaugino masses, otherwise the posterior pdfs resulting from the two scans agree well in all three planes. This is encouraging, since it indicates that experimental data is becoming constraining enough to dominate the inference results, which are now only weakly dependent on the choice of prior, although some residual volume effects remain in the relative posterior weight of the FP region with respect to other parts of the parameter space.

The 2D profile likelihood results\footnote{In order to construct the profile likelihood contours presented here, it was necessary to remove from our scans the highest likelihood point, which showed an anomalously low $\chi^2$. This could be traced back to the fact that this point had a very small mass difference between the second-heaviest neutralino and the heaviest smuon, with a fractional difference of less than $10^{-6}$. This led to reproducing very closely the observed anomalous magnetic moment of the muon, but with a large amount of numerical fine tuning. We removed this finely-tuned point from the scan.}  are shown in the bottom panels of Fig.~\ref{fig:ALL_NoXe_LHC}. Similarly to the Bayesian analysis, the LHC 2011 exclusion limit cuts deep into the stau-coannihilation region. The favoured region of parameter space is still located at relatively small scalar and gaugino masses, not far above the new LHC exclusion limit. The profile likelihood results are qualitatively very similar to the Bayesian results, most importantly the FP region is still allowed at $99\%$ confidence level. The $99\%$ contours for the profile likelihood are very similar to the extent of the $99\%$ Bayesian credible intervals shown in the upper panels. As we already pointed out in Ref.~\cite{arXiv:1107.1715}, the $99\%$ confidence region is much larger than expected from the $68\%$ and $95\%$ contours, indicating that the tails of the profile likelihood function are highly non-Gaussian. This demonstrates that the use of a high resolution scan such as the one we employ in this paper is required in order to correctly map out the shape of the profile likelihood function and accurately infer the extent of the 99$\%$ confidence level.

We now examine the best fit points in more detail. Details of the best fit points are given in Table~\ref{tab:bestfit}, a breakdown of the total $\chi^2$ by observable is given in Table~\ref{tab:chisquare}, where $\chi^2$ can be found by summing over the likelihood $\mathcal {L}_i$ for each observable
\begin{equation}
\chi^2 \equiv -2 \sum_{i} \ln {\mathcal L}_i .
\end{equation}
For observables that are constrained by exclusion limits we choose $\ln {\mathcal L}_i = 0$ for points outside the allowed region. An exception is the XENON100 exclusion limit, as described in detail in Ref.~\cite{arXiv:1107.1715}. For Gaussian-distributed observables we normalize the likelihood such that $\ln {\mathcal L}_i = 0$ when the theoretical and the experimental values match. In this analysis we use 13 ``active'' Gaussian data points (see Table~3 in Ref.~\cite{arXiv:1107.1715}) and scan over 4 free cMSSM parameters. Therefore, in total there are 9 degrees of freedom.
 
The best fit point for this scan is found in the SC region, corresponding to small scalar and gaugino masses $< 1$ TeV that are still allowed by LHC 2011 constraints.
The observables having the largest contribution to the total  $\chi^2$ in Table~\ref{tab:chisquare} are the isospin asymmetry $\DeltaO$ and the branching ratios of the $\Dstaunu$ and $\Dsmunu$ 
processes which are sensitive to new physics mainly through the mass of the 
charged Higgs bosons $H^\pm$ and to some extend to $\tan\beta$~\cite{Akeroyd:2009tn}, in particular at low $m_{H^\pm}$. 
Our best fit points have relatively large $m_{H^\pm}$ and therefore the 
theoretical values for those processes are SM-like and have a discrepancy at 
the $ \sim 1.5 \, \sigma$ level with the experimental values.
For the isospin asymmetry $\DeltaO$ new physics contributions 
become important mainly to low $m_{1/2}$ values~\cite{Ahmady:2006yr}. 
Hence the predictions for our best fit points are again SM-like, showing a 
discrepancy at the $2 \, \sigma$ level with the experimental central values.


\begin{figure*}
\centering
\includegraphics[width=0.32\linewidth]{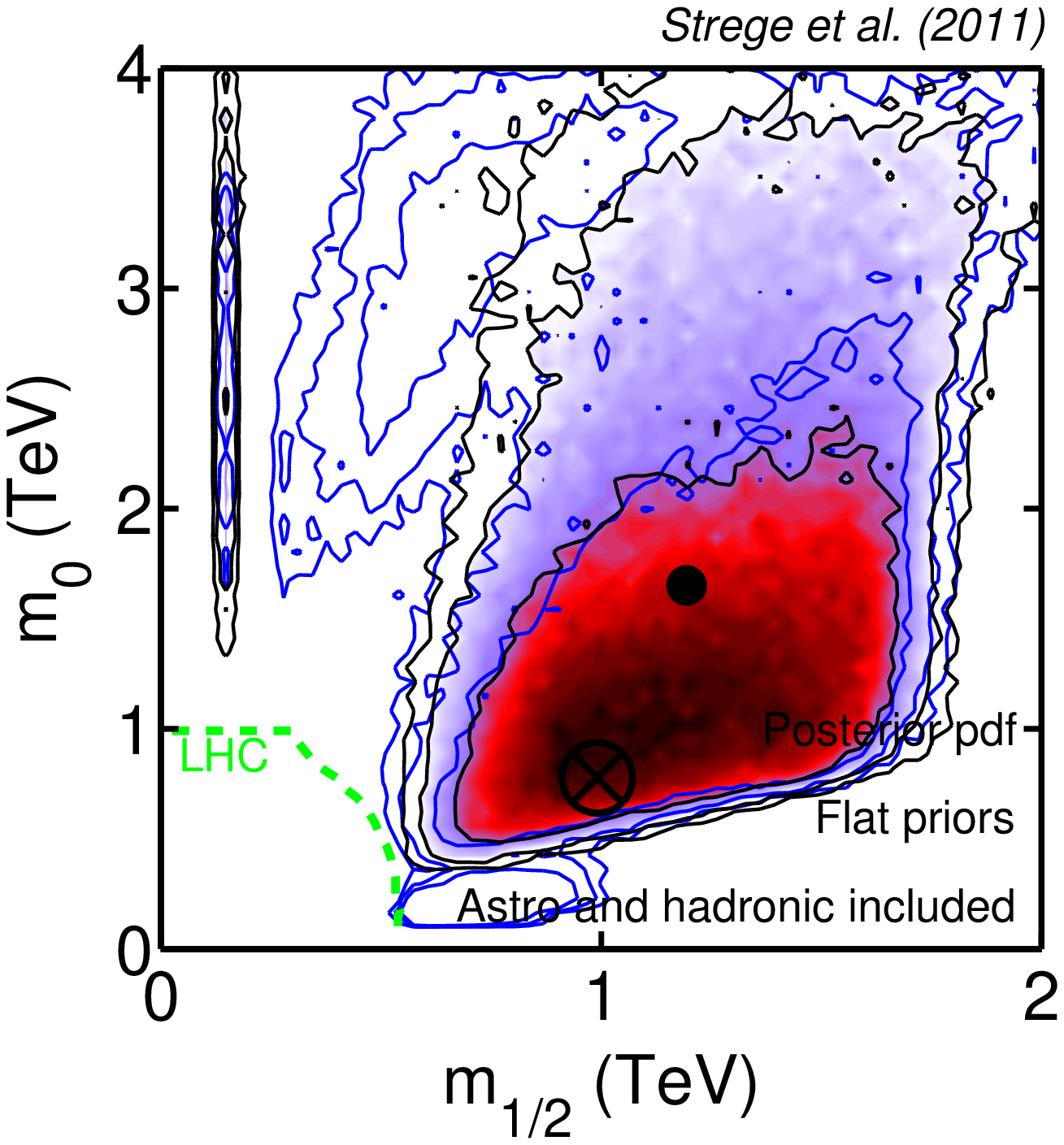}
\includegraphics[width=0.32\linewidth]{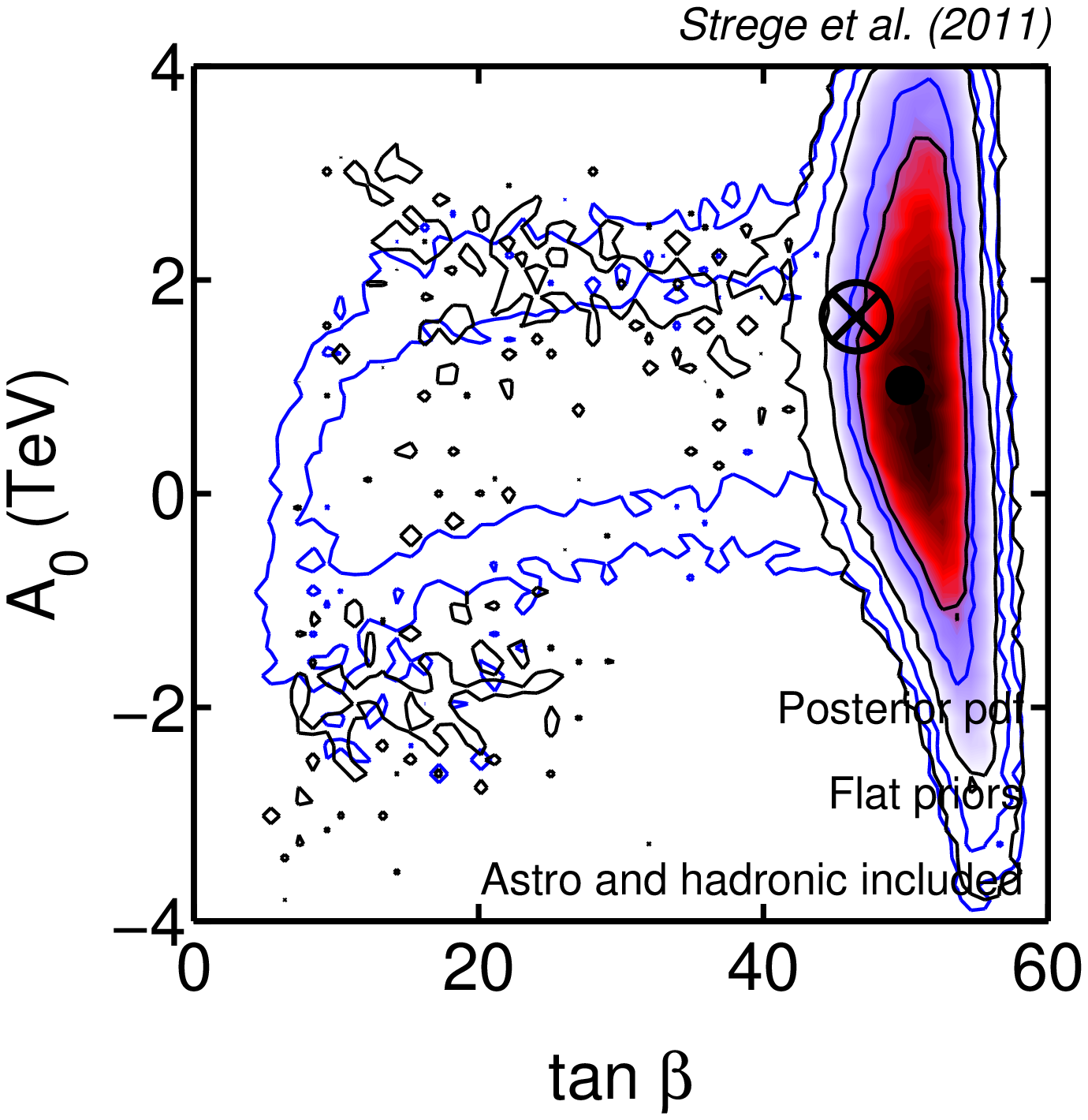}
\includegraphics[width=0.32\linewidth]{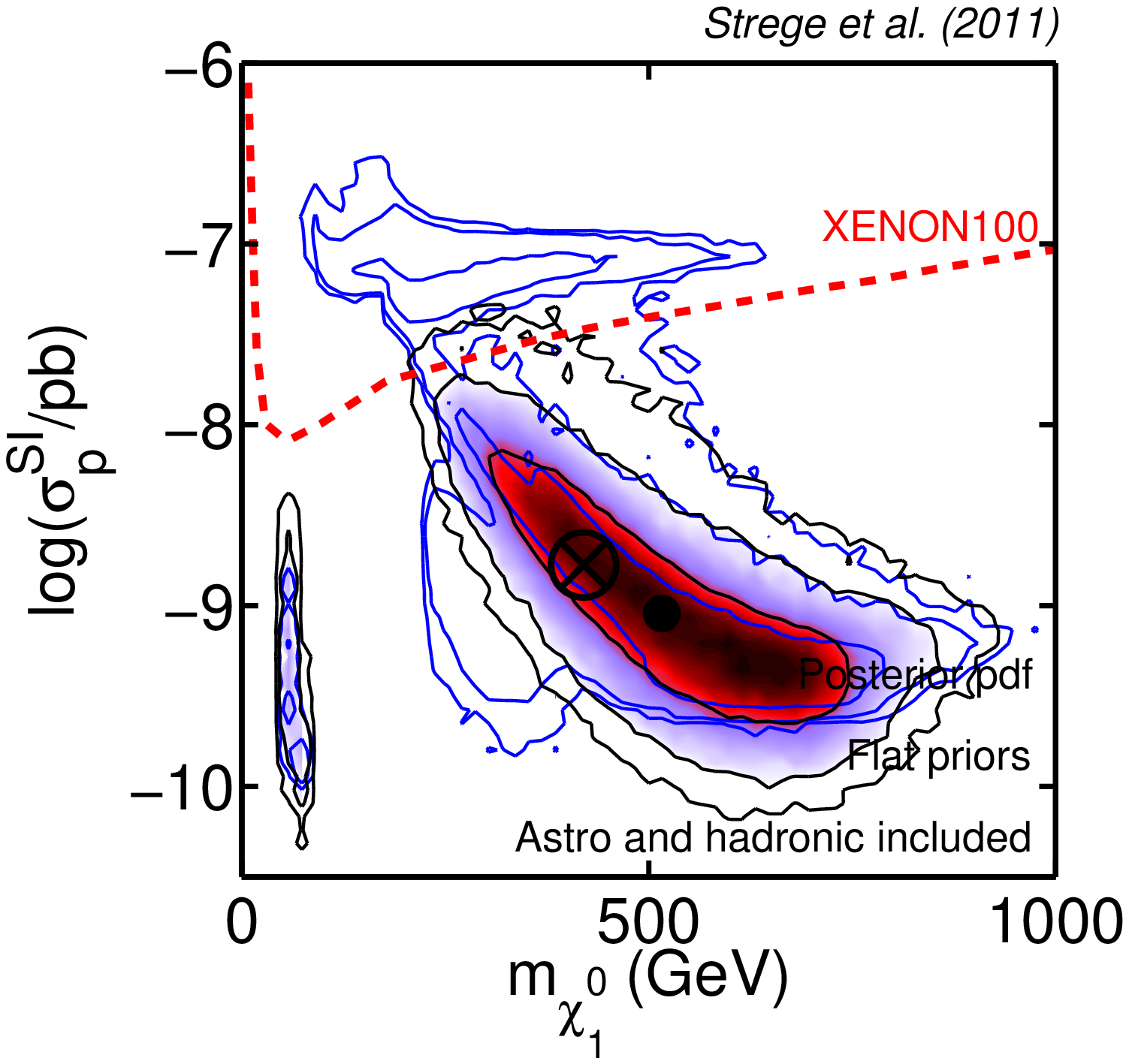}
\includegraphics[width=0.32\linewidth]{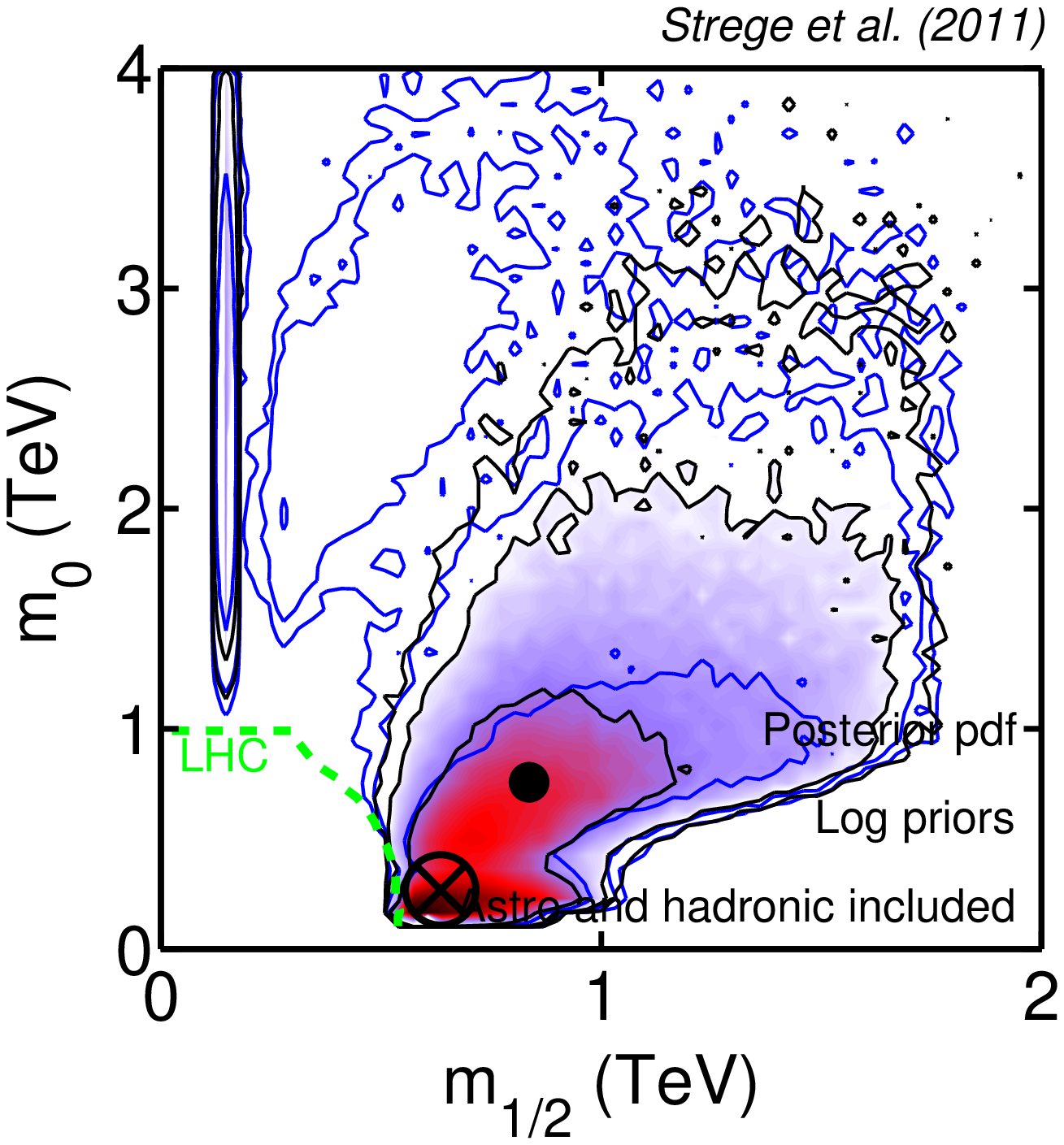}
\includegraphics[width=0.32\linewidth]{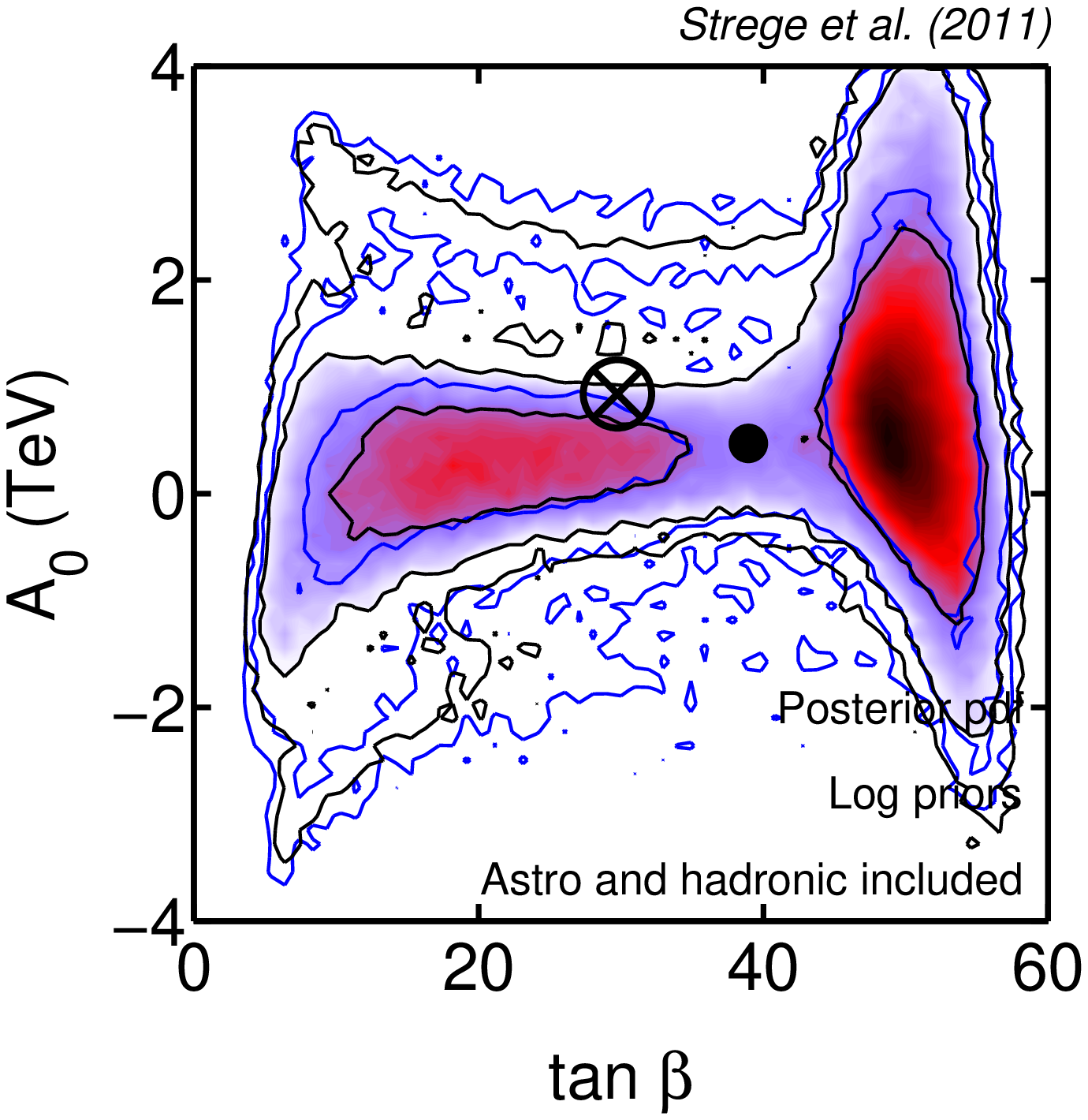}
\includegraphics[width=0.32\linewidth]{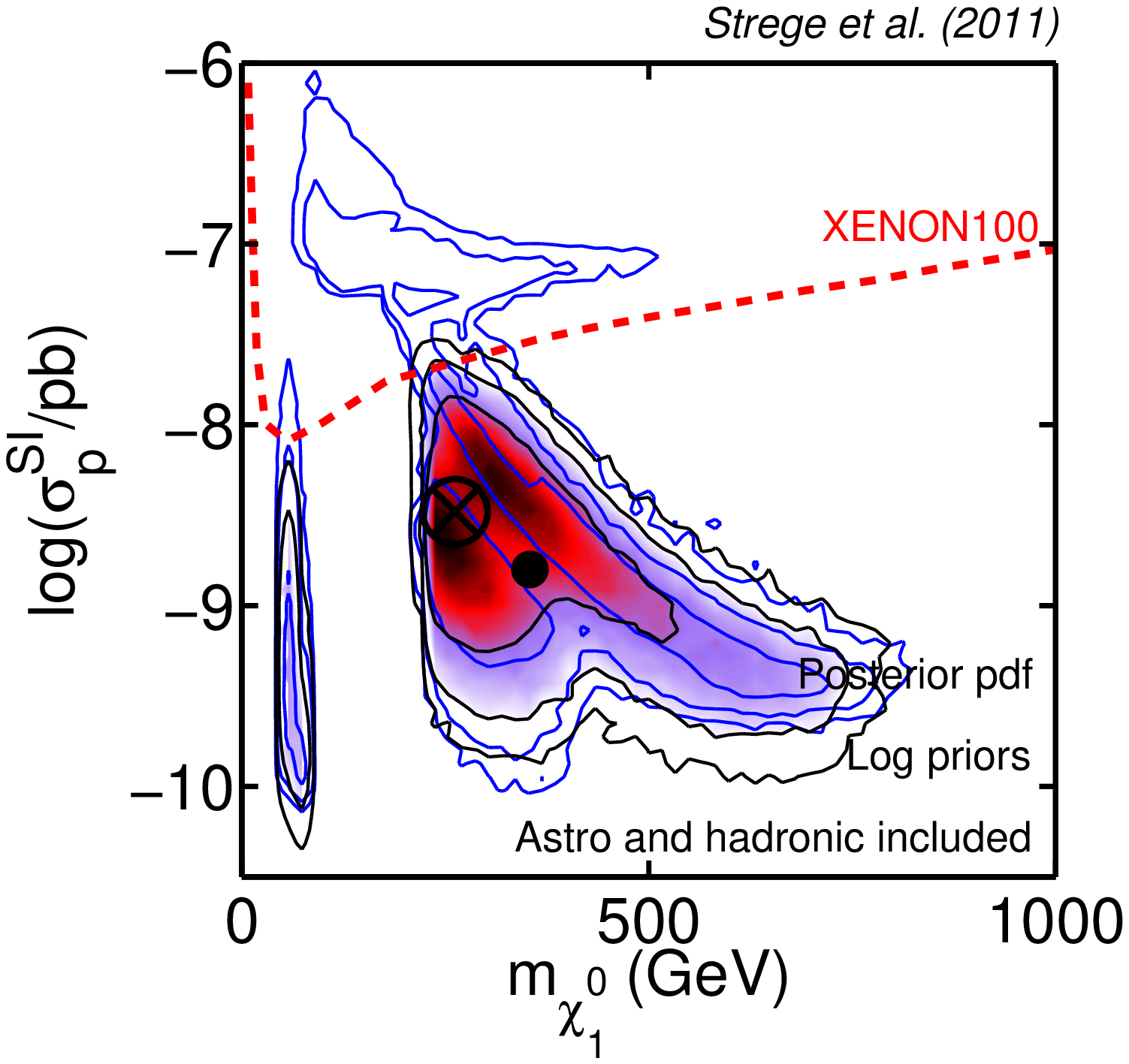}
\includegraphics[width=0.32\linewidth]{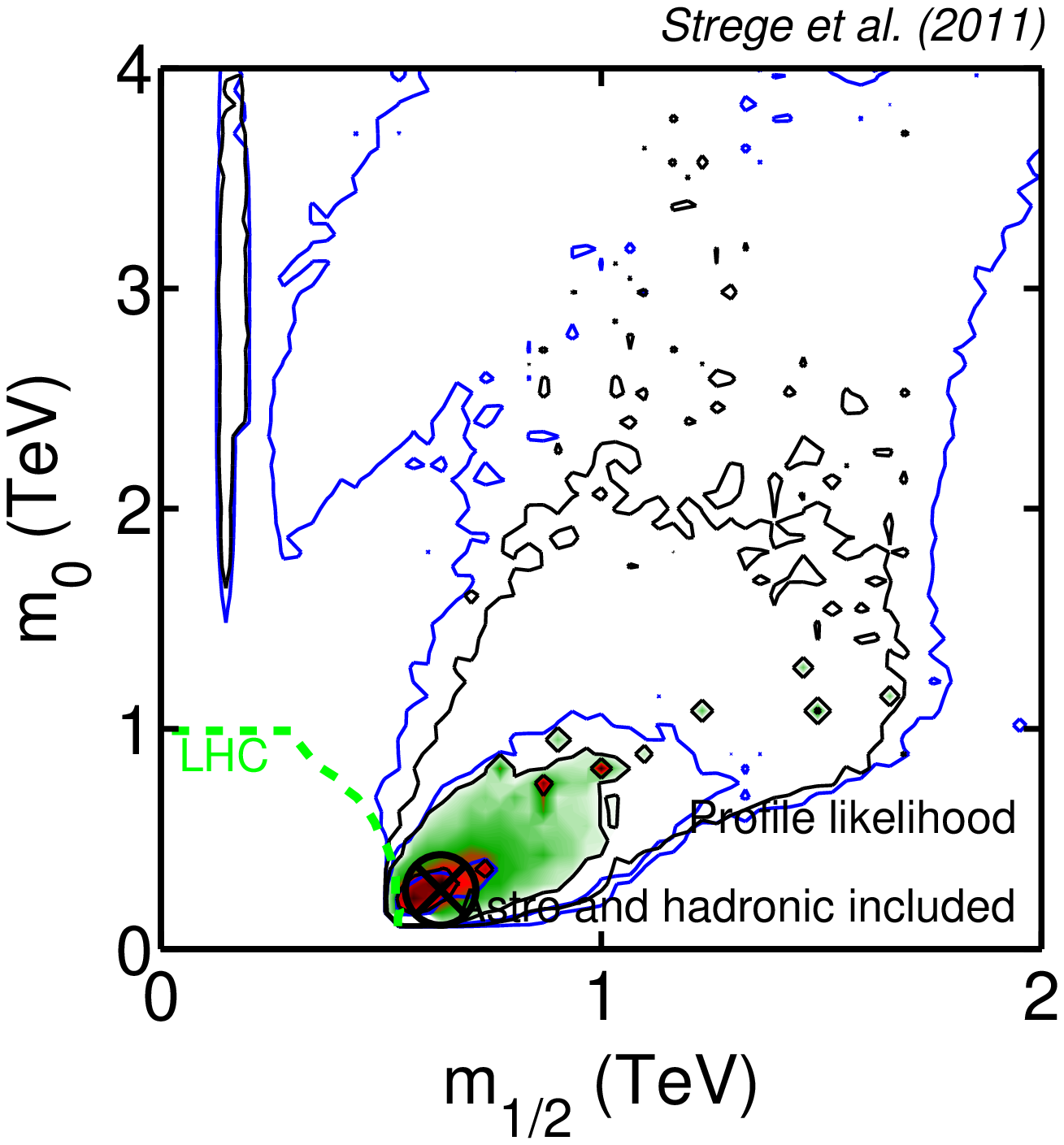}
\includegraphics[width=0.32\linewidth]{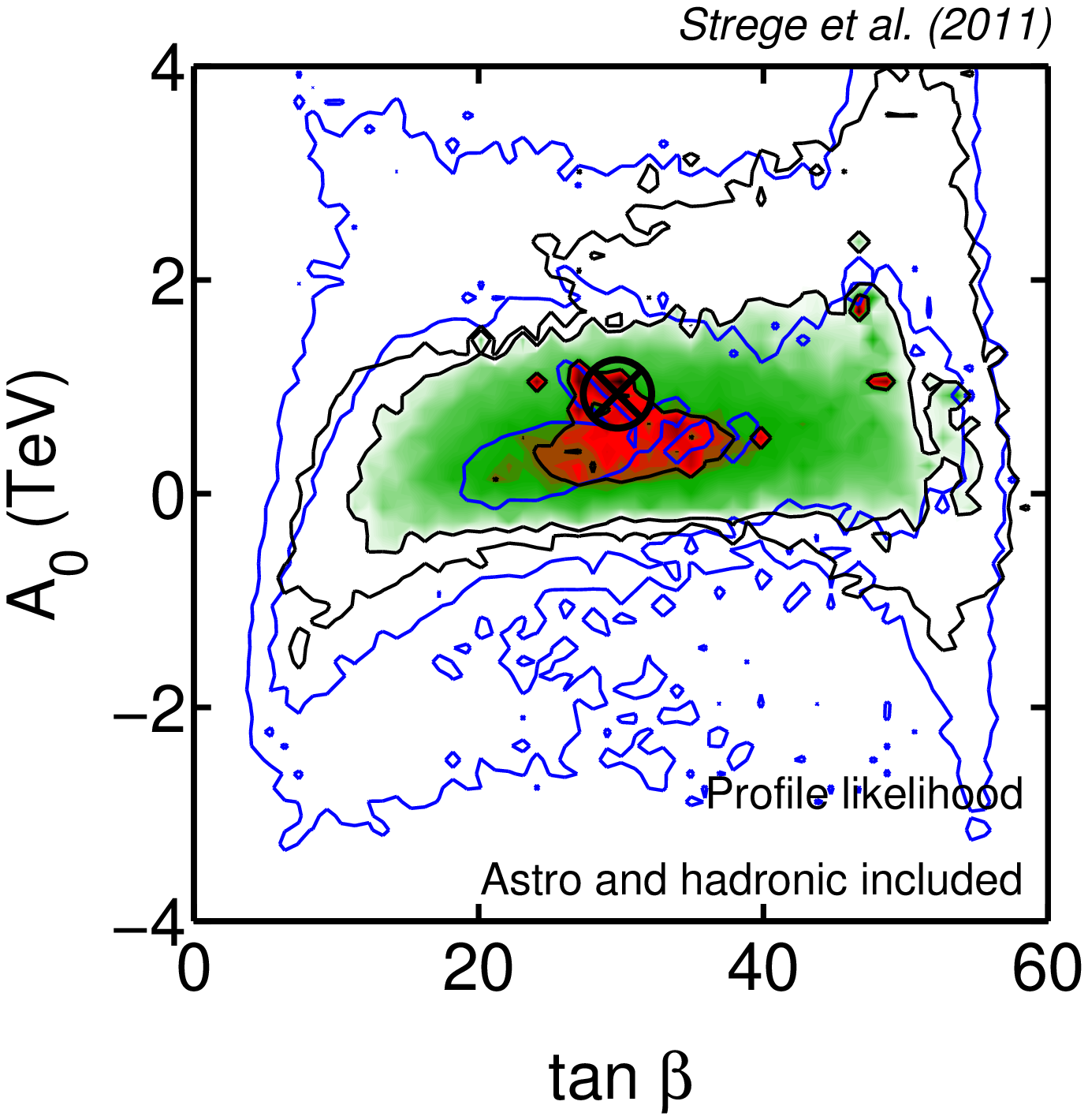}
\includegraphics[width=0.32\linewidth]{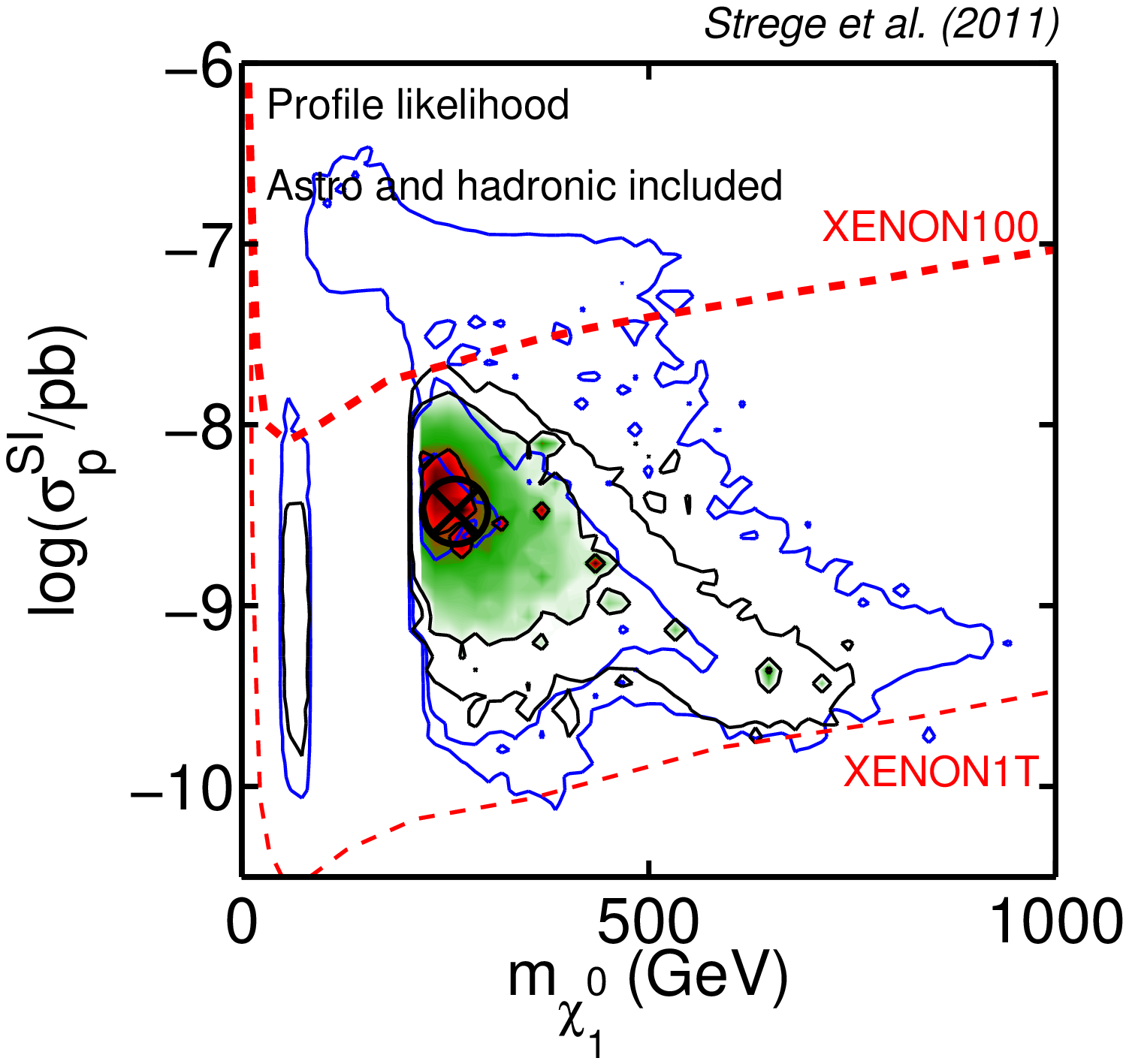}
\caption{\fontsize{9}{9} \selectfont As in Fig.~\ref{fig:ALL_NoXe_LHC}, but now black contours include XENON100 data (considering hadronic and astrophysical uncertainties as nuisance parameters), while the blue empty contours show for comparison the case where no direct detection data are included (from the inside out: 68\%, 95\% and 99\% regions). We observe a strong suppression of the viability of the FP region. Notice that the XENON100 90\% limit (red/dashed line) has been included only to guide the eye, as our implementation of the XENON100 data is slightly more conservative than the procedure adopted in Ref.~\cite{xenon:2011hi}. \label{fig:ALL_Xe_LHC}
}
\end{figure*}

\subsection{Impact of XENON100 data, including hadronic and astrophysical uncertainties}

The effect of adding the XENON100 exclusion limit is shown in  Fig.~\ref{fig:ALL_Xe_LHC}. Astrophysical and hadronic uncertainties are accounted for by marginalising/profiling over the corresponding nuisance parameters. We have checked that, in agreement with the findings of Ref.~\cite{arXiv:1107.1715}, the overall impact of including these uncertainties in the analysis leaves our results qualitatively unchanged, though quantitatively it affects the overall limits. 

As can be seen in the upper two panels of Fig.~\ref{fig:ALL_Xe_LHC}, the inclusion of XENON100 data has a significant impact on the two-dimensional posterior distributions for the cMSSM. The focus point region, which could previously not be excluded at the $99\%$ level, is now strongly disfavoured; the parameter space is significantly reduced. Aside from residual volume effects the contours in the $(m_{\neut},\sigmaSI)$ plane for the two different choices of priors are in relatively good agreement with each other. Note that for the posterior with flat priors the SC region is strongly disfavoured. This is not a physical effect, as can be seen from the fact that the combined best fit point from the flat and log prior scans is found inside the SC region. Instead, this is a result of an underexploration of the cMSSM parameter space by the flat prior scan. Due to the high dimensionality of the parameter space small regions of high likelihood can easily be missed. The flat prior scan is particularly vulnerable to this, since it explores the low mass regions in much less detail than the log prior scan.

By comparing the top and central right-hand panels in Fig.~\ref{fig:ALL_Xe_LHC} we can see that the 99$\%$ contours are very similar for the posterior pdfs for both choices of priors. Thus, the constraints on $m_{\neut}$ and $\sigmaSI$ are becoming increasingly independent of the choice of priors. 

The PL results, shown in the bottom panels, qualitatively agree well with the Bayesian analysis. The parameter space shrinked significantly. The contours are tighter than the Bayesian credible regions, which is explained by the high likelihood value of the best fit point (see table~\ref{tab:chisquare}). This illustrates that profile likelihood results have to be interpreted with care; the extent of the contours strongly depends on the likelihood value of a single point in parameter space found by the scan. Again the best fit point is found in the SC region. Note that this best fit point corresponds to a relatively low $m_h$, such that a sizeable contribution to the $\chi^2$ arises from the 5 fb$^{-1}$ $95\%$ Higgs exclusion limit. Similar to the results of the Bayesian analysis, the FP region is ruled out at 99$\%$ confidence level by the XENON100 data, as can be seen explicitly in the rightmost plot. This clearly illustrates the potential of direct detection experiments to constrain SUSY.

\subsection{Impact of the  $\delta a_\mu^{SUSY}$ constraint}

\begin{figure*}
\centering
\includegraphics[width=0.32\linewidth]{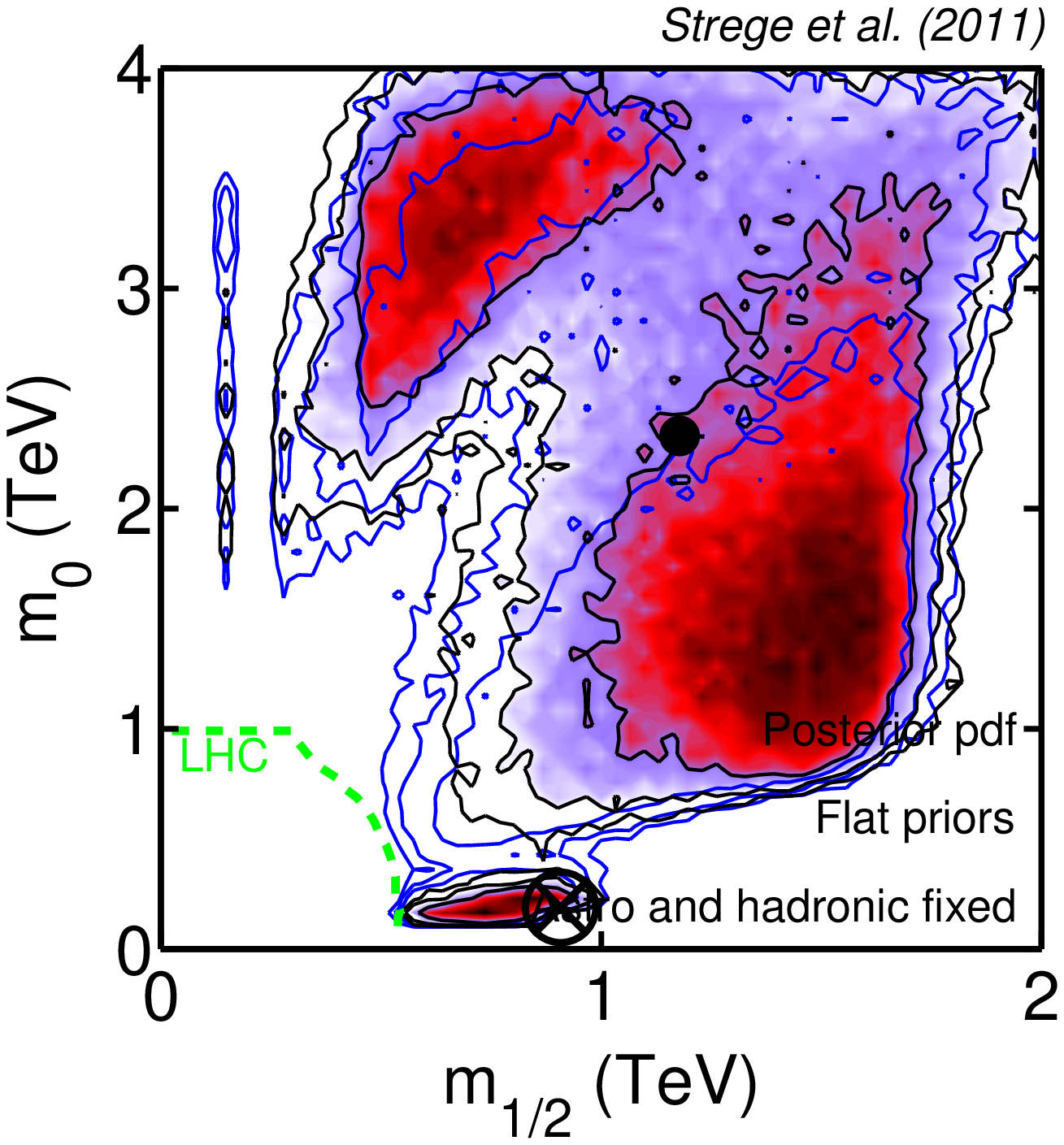}
\includegraphics[width=0.32\linewidth]{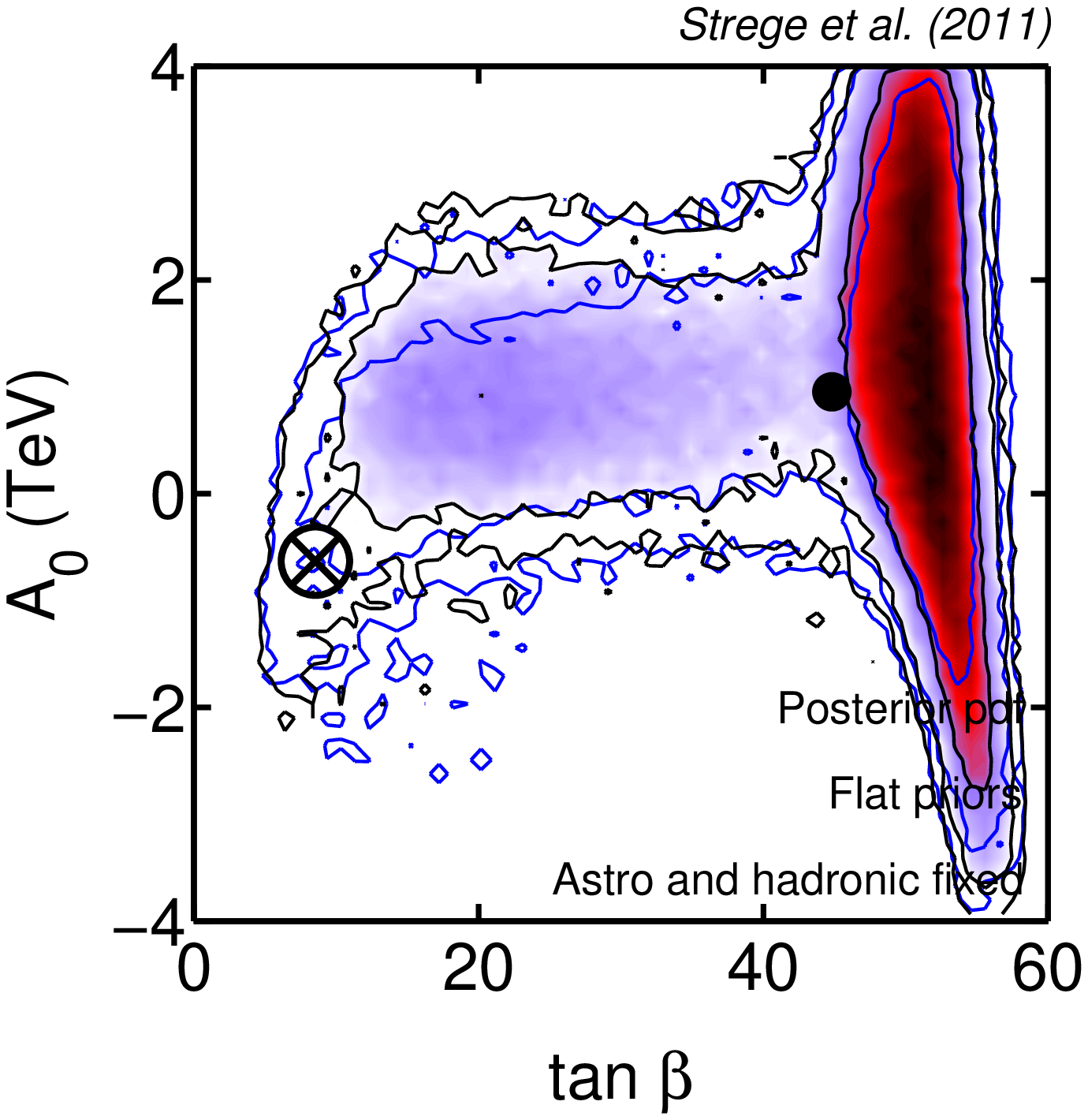}
\includegraphics[width=0.32\linewidth]{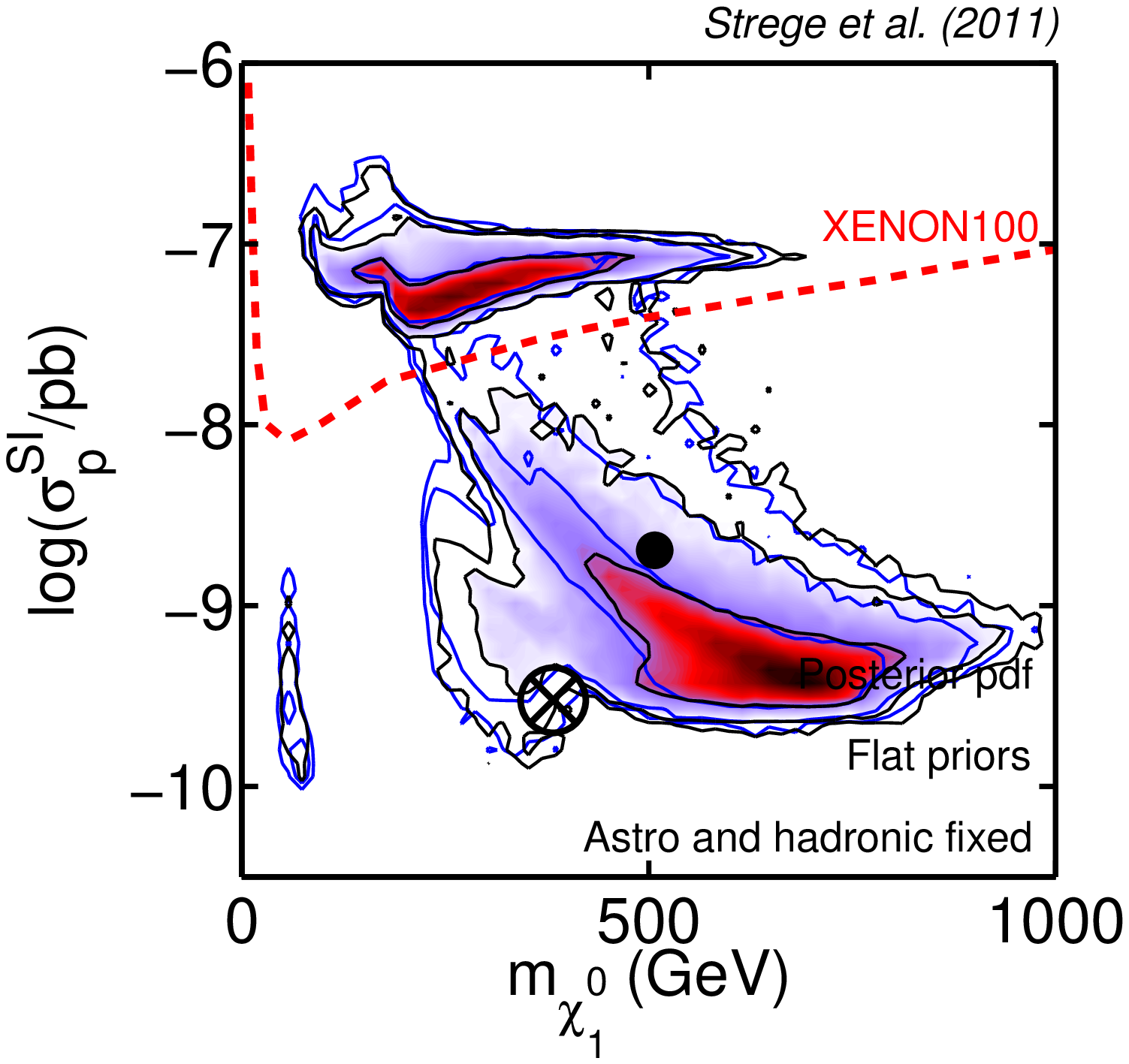} \\
\includegraphics[width=0.32\linewidth]{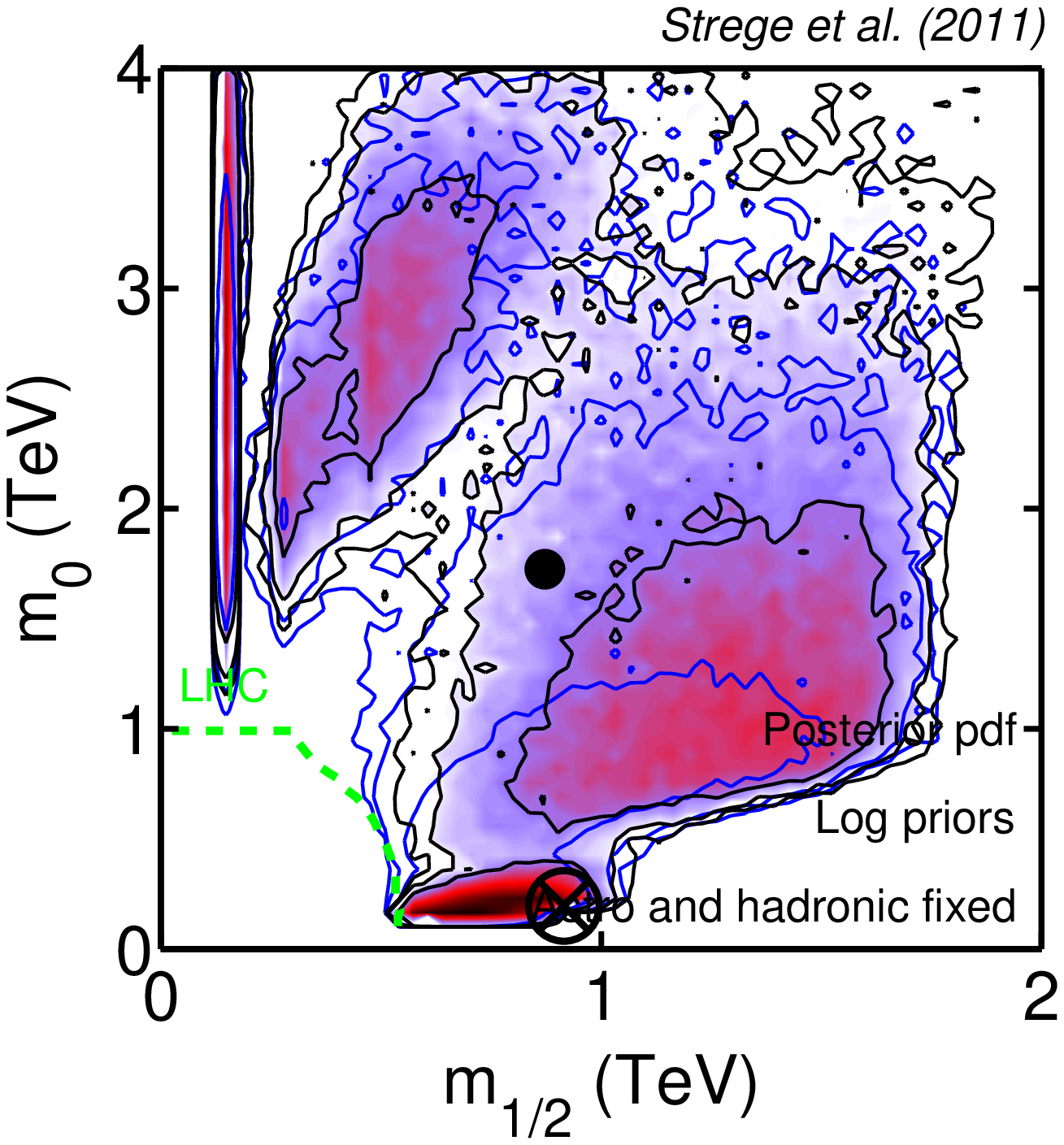}
\includegraphics[width=0.32\linewidth]{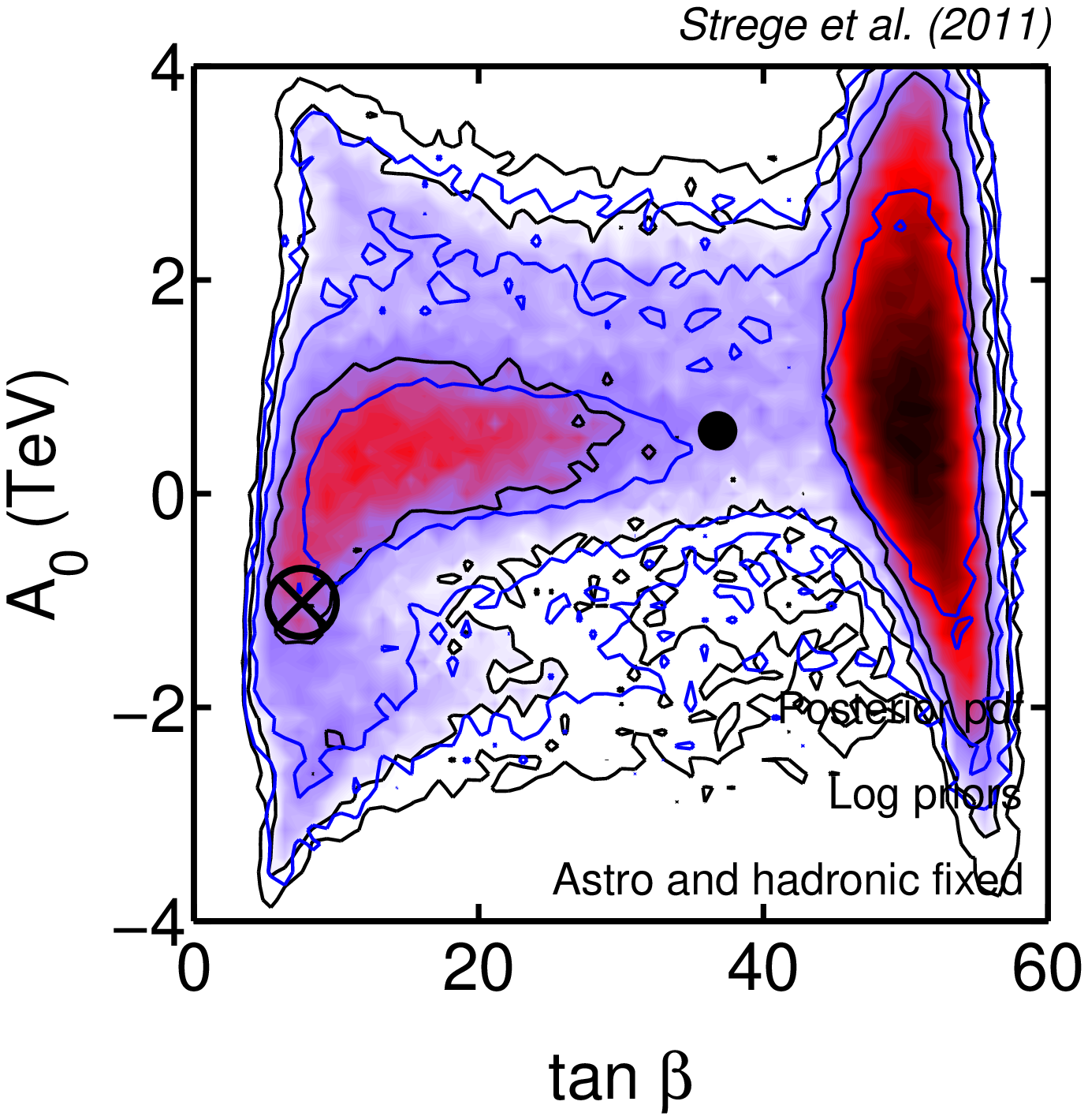}
\includegraphics[width=0.32\linewidth]{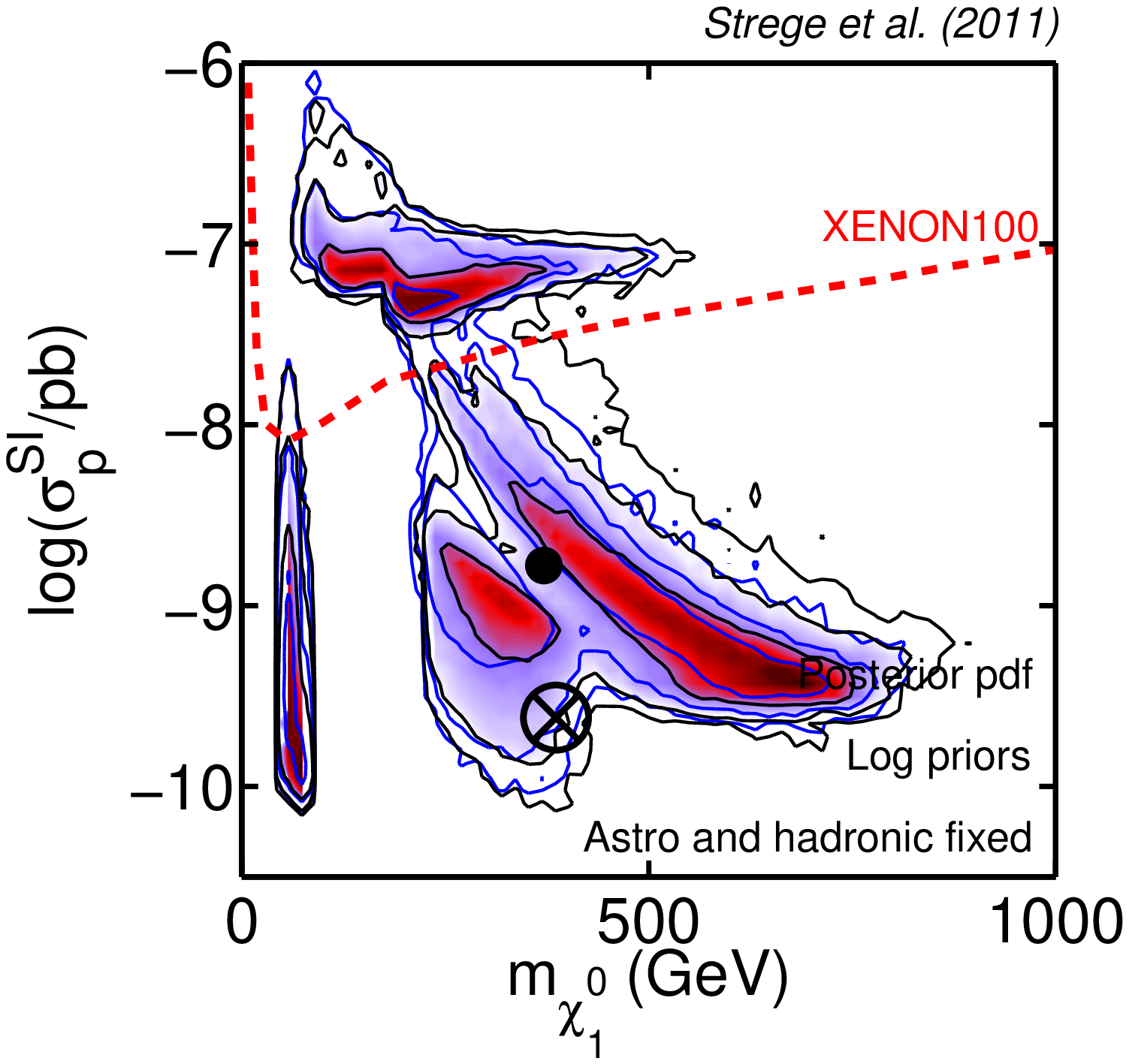}
\includegraphics[width=0.32\linewidth]{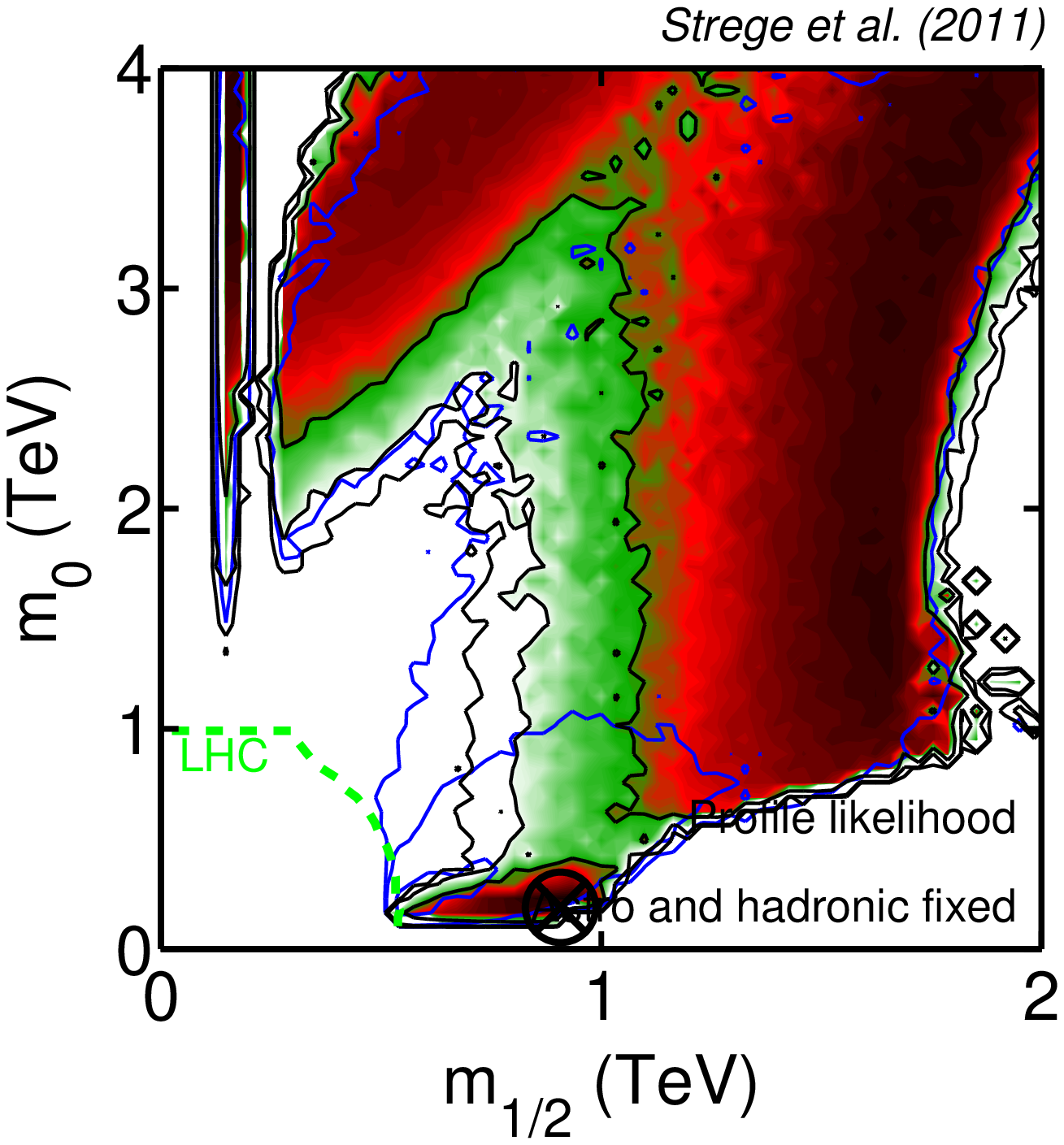}
\includegraphics[width=0.32\linewidth]{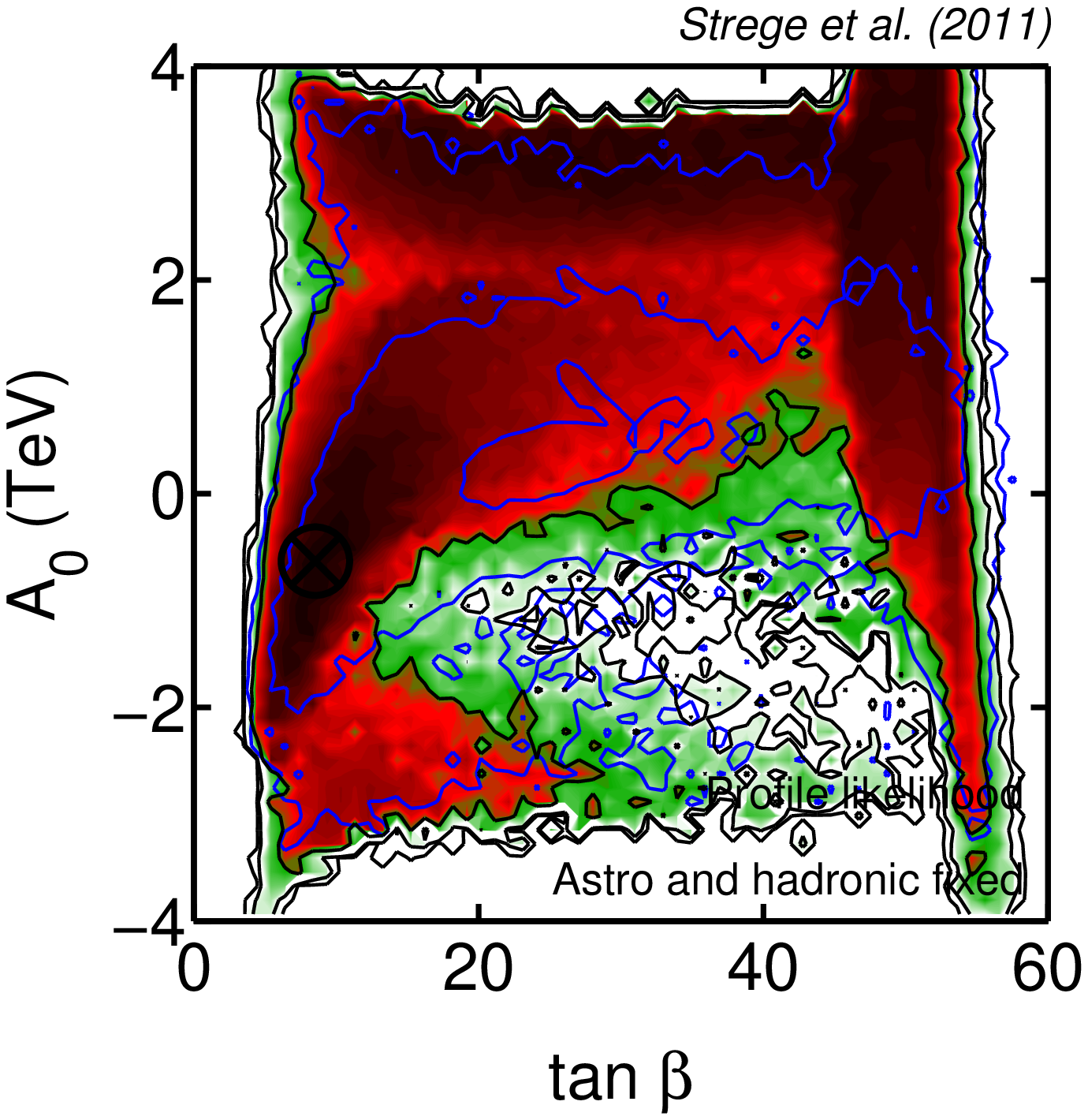}
\includegraphics[width=0.32\linewidth]{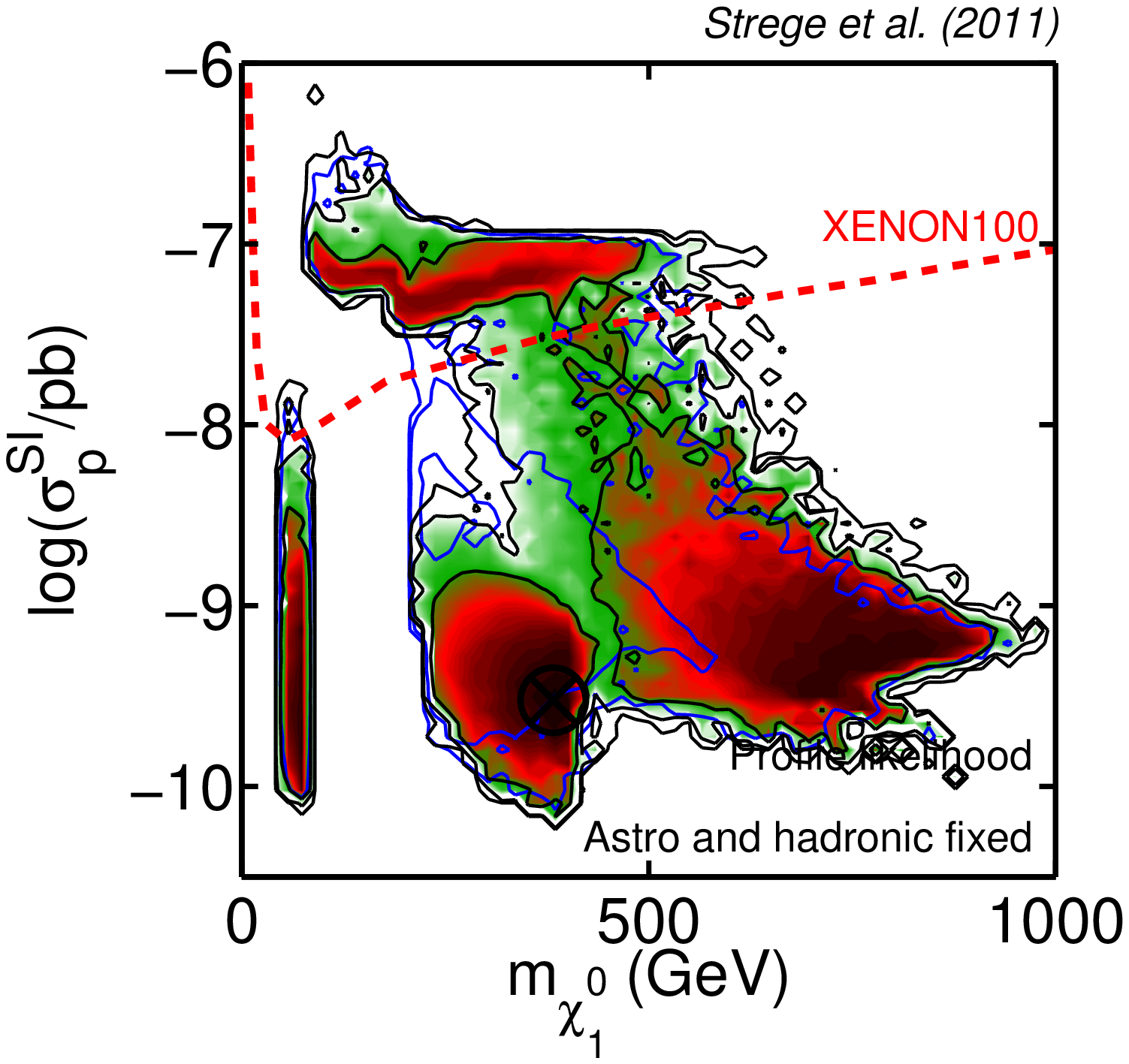}
\caption{\fontsize{9}{9} \selectfont As in Fig.~\ref{fig:ALL_NoXe_LHC}, but now black/filled contours result from scans that do not include the experimental constraint on the anomalous magnetic moment of the muon. Blue contours correspond to the results obtained when applying this constraint, and thus are identical to the black contours in Fig.~\ref{fig:ALL_NoXe_LHC}. \label{fig:no_gm2}}
\end{figure*}

The muon anomalous magnetic moment provides an interesting window to new Physics, since it is very accurately measured. A constraint on the
supersymmetric contribution to this observable, $\delta a_\mu^{SUSY}$, can be
extracted by comparing the experimental result
\cite{g-2}, with the theoretical evaluations of the
Standard Model contribution \cite{Jegerlehner:2009ry,Davier:2010nc,Hagiwara:2011af}.
Although the latter have become increasingly precise in the last decade, they are still subject to theoretical uncertainties, especially in the computation of the hadronic loop contributions.
According to the most recent evaluations, when $e^+e^-$ data are used the experimental excess in
$a_\mu\equiv(g_\mu-2)/2$ would constrain a possible supersymmetric
contribution to be $\delta a_\mu^{SUSY}=(29.6\pm8.1)\times10^{-10}$, where
theoretical and experimental errors have been combined in
quadrature. However, when tau data are used a smaller discrepancy ($2.4\,\sigma$)
with the experimental measurement is found \cite{Davier:2010nc}.

In Ref.~\cite{Trotta:2008bp} it has been shown that the preference for small $m_0$ and $m_{1/2}$ in global fits of the cMSSM is strongly driven by the $\delta a_\mu^{SUSY}$ constraint. In order to evaluate the dependence of our results on this constraint, we repeat the analysis presented in Section~\ref{sec:LHC_only} after dropping the experimental constraint on  $\delta a_\mu^{SUSY}$. The results are shown in Fig.~\ref{fig:no_gm2}. For comparison, blue contours show the constraints derived on the cMSSM when including the $\delta a_\mu^{SUSY}$ constraint in the analysis.

As can be seen the posterior contours for both sets of priors are significantly expanded outwards towards larger values of $m_0$ and $m_{1/2}$. The FP region is allowed at the $68\%$ level. Clearly, the $\delta a_\mu^{SUSY}$ constraint is the dominating factor for the exclusion of high scalar masses $m_0 \sim$ a few TeV. No other constraints strongly disfavour these regions. However, even when excluding the $\delta a_\mu^{SUSY}$ constraint from the scan, the posterior probability for small scalar and gaugino masses remains high; these mass scales are favoured by a number of different constraints. 

These effects are even more pronounced for the profile likelihood analysis (bottom panels of Fig.~\ref{fig:no_gm2}). The extent of all three contours increases significantly when the $\delta a_\mu^{SUSY}$ constraint is excluded from the scans. High scalar and gaugino masses are allowed at the $68\%$ level. Without the constraint on the anomalous magnetic moment of the muon no conclusions can be drawn on the allowed ranges of $\tan \beta$ and $A_0$. The allowed region of parameter space in the $(m_{\neut},\sigmaSI)$ plane also increases significantly. The $\chi^2$/dof for the best fit point, shown in Table~\ref{tab:chisquare}, is significantly reduced, which emphasises that the $\delta a_\mu^{SUSY}$ constraint is in strong conflict with several other constraints. However, the best fit point is still found at small $m_0$ and $m_{1/2}$, such that the SC region remains the most favoured region even when dropping the $\delta a_\mu^{SUSY}$ constraint.

Therefore, global fits of the cMSSM are still strongly driven by the $\delta a_\mu^{SUSY}$ constraint, which significantly disfavours large scalar and gaugino masses and therefore pushes both the posterior and the profile likelihood contours to small values of $m_0$ and $m_{1/2}$.

\section{Updated prospects for cMSSM discovery and Higgs excess} 
\label{sec:prospects}

\begin{figure*}
\centering
\includegraphics[width=0.242\linewidth]{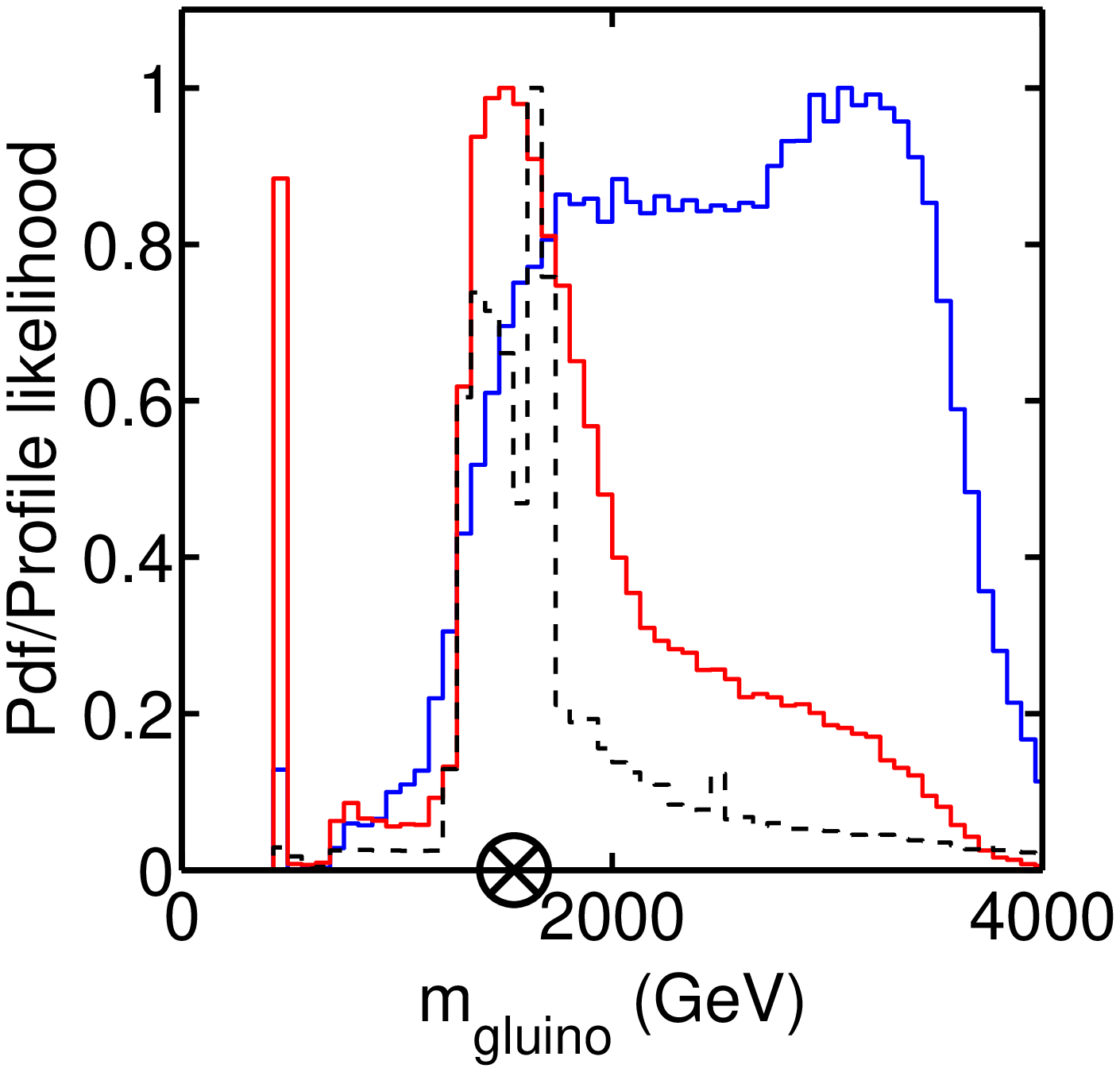}
\includegraphics[width=0.242\linewidth]{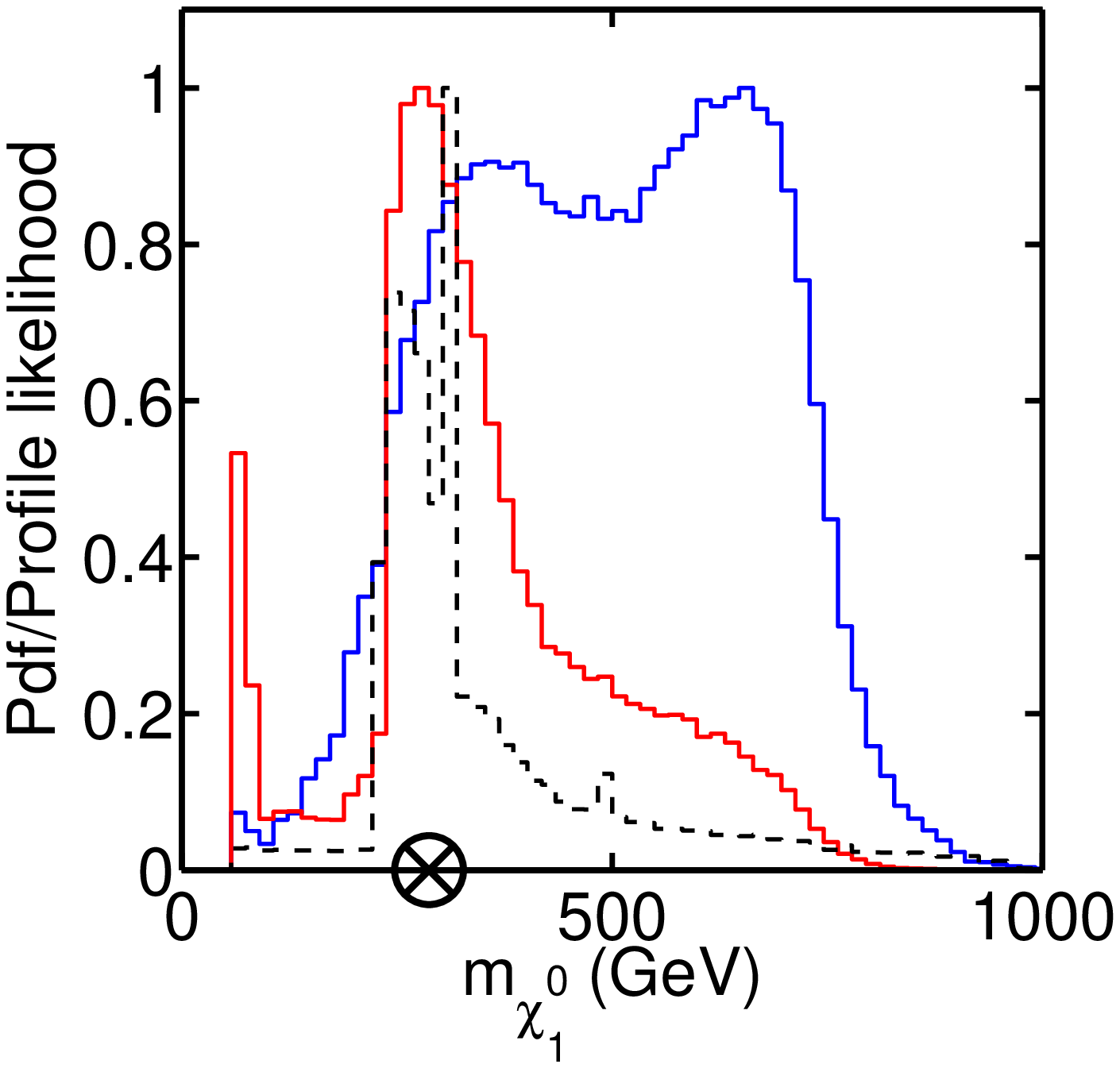}
\includegraphics[width=0.242\linewidth]{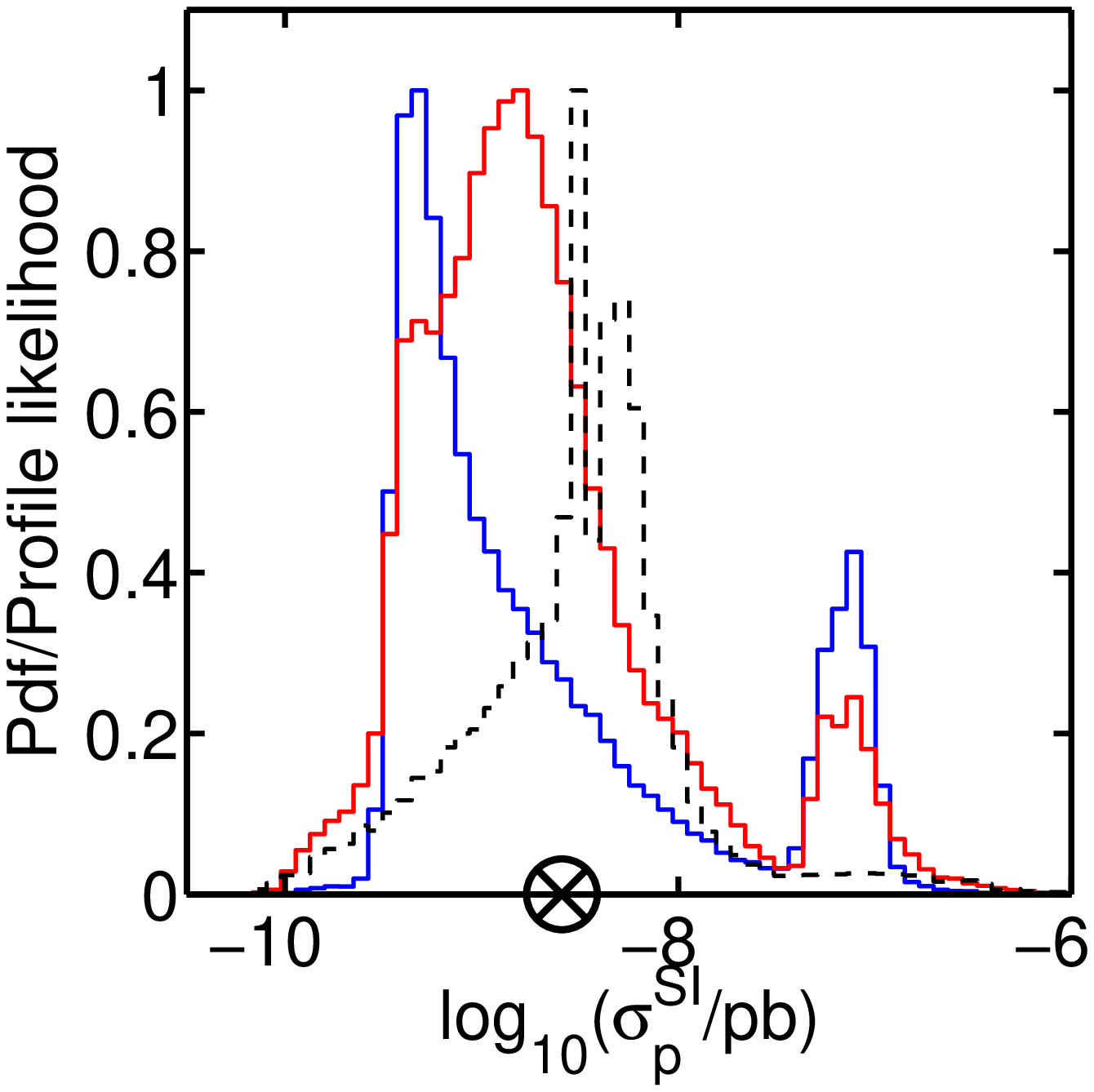}
\includegraphics[width=0.242\linewidth]{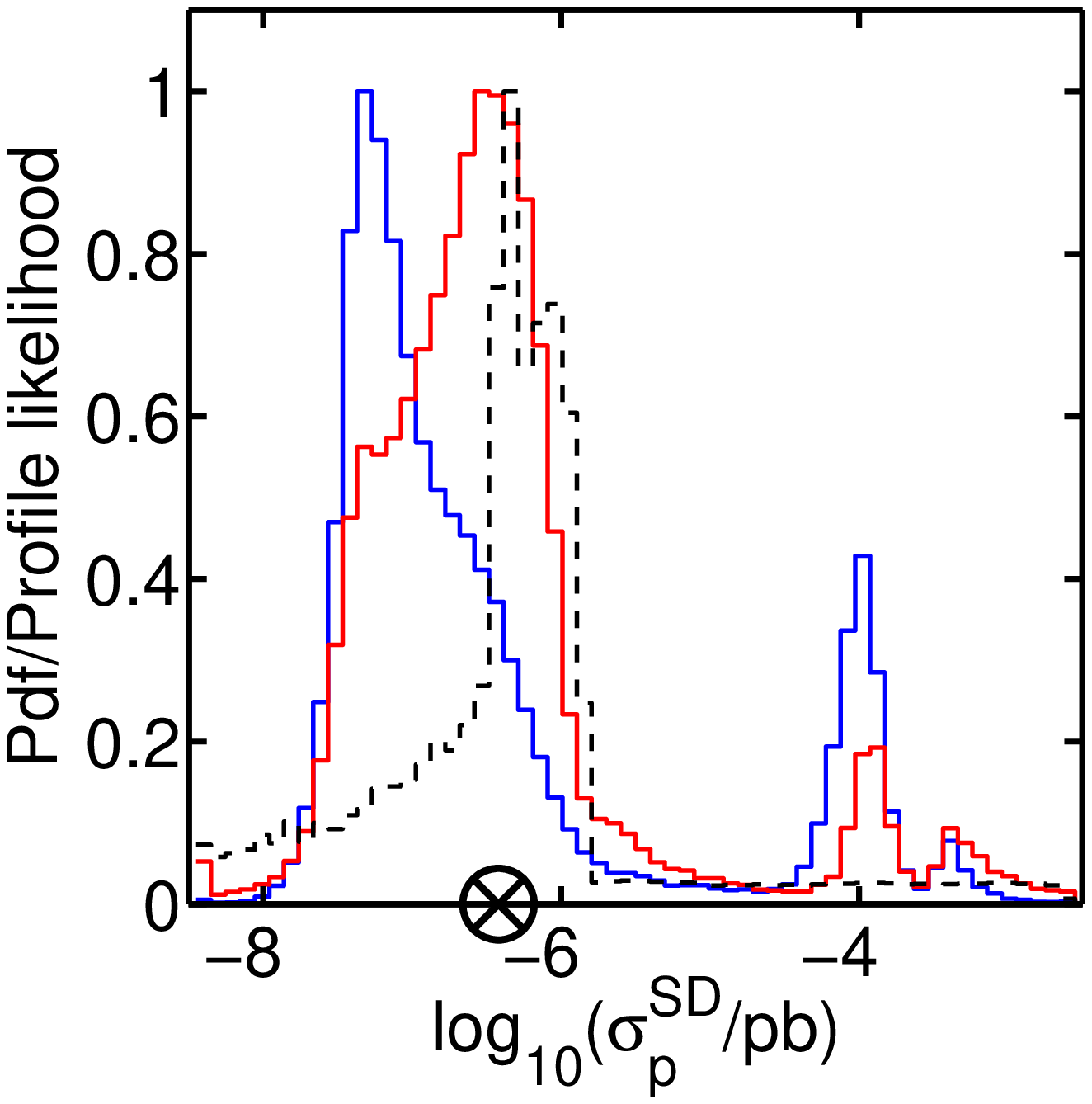} \\ 
\includegraphics[width=0.242\linewidth]{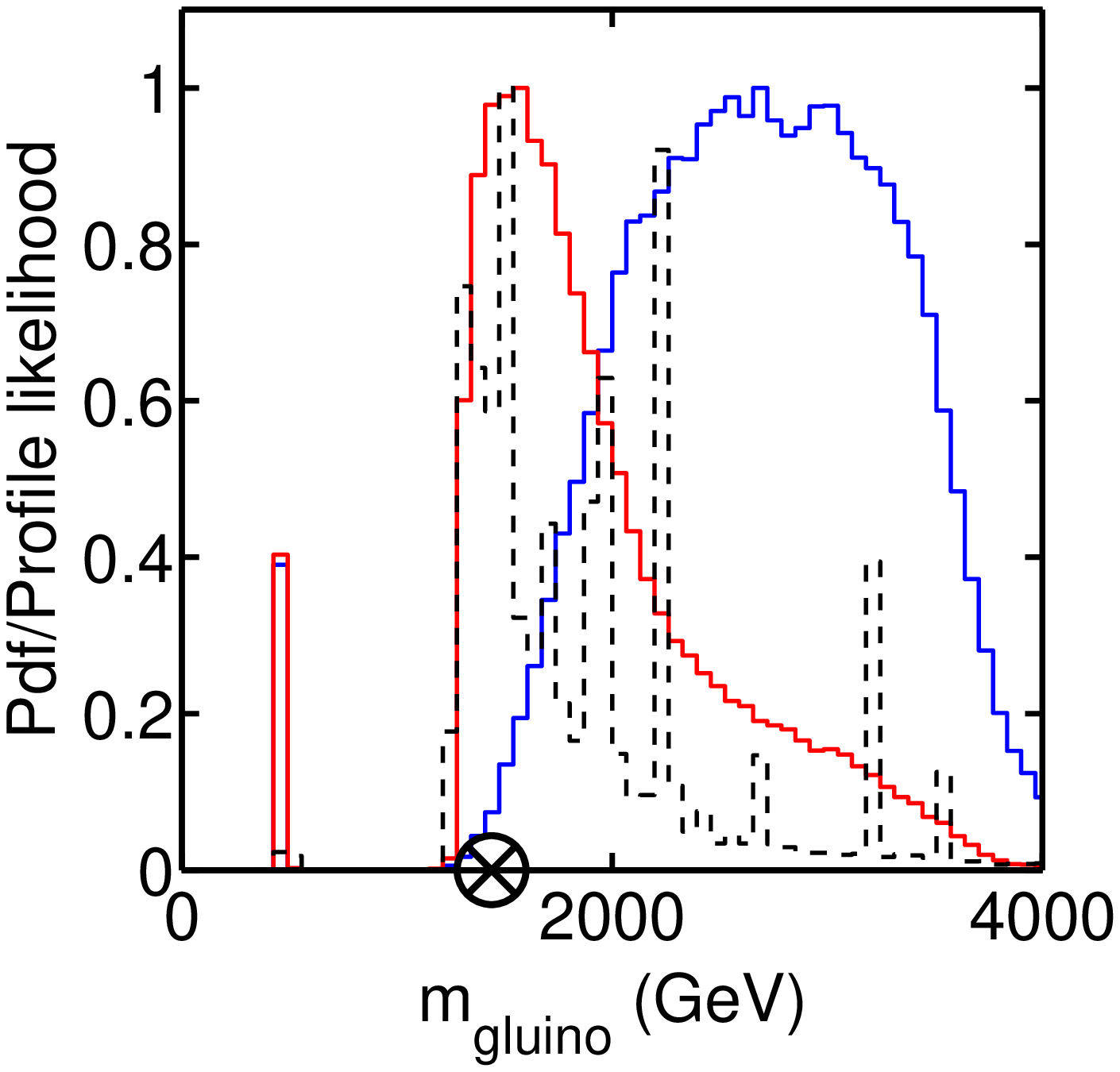}
\includegraphics[width=0.242\linewidth]{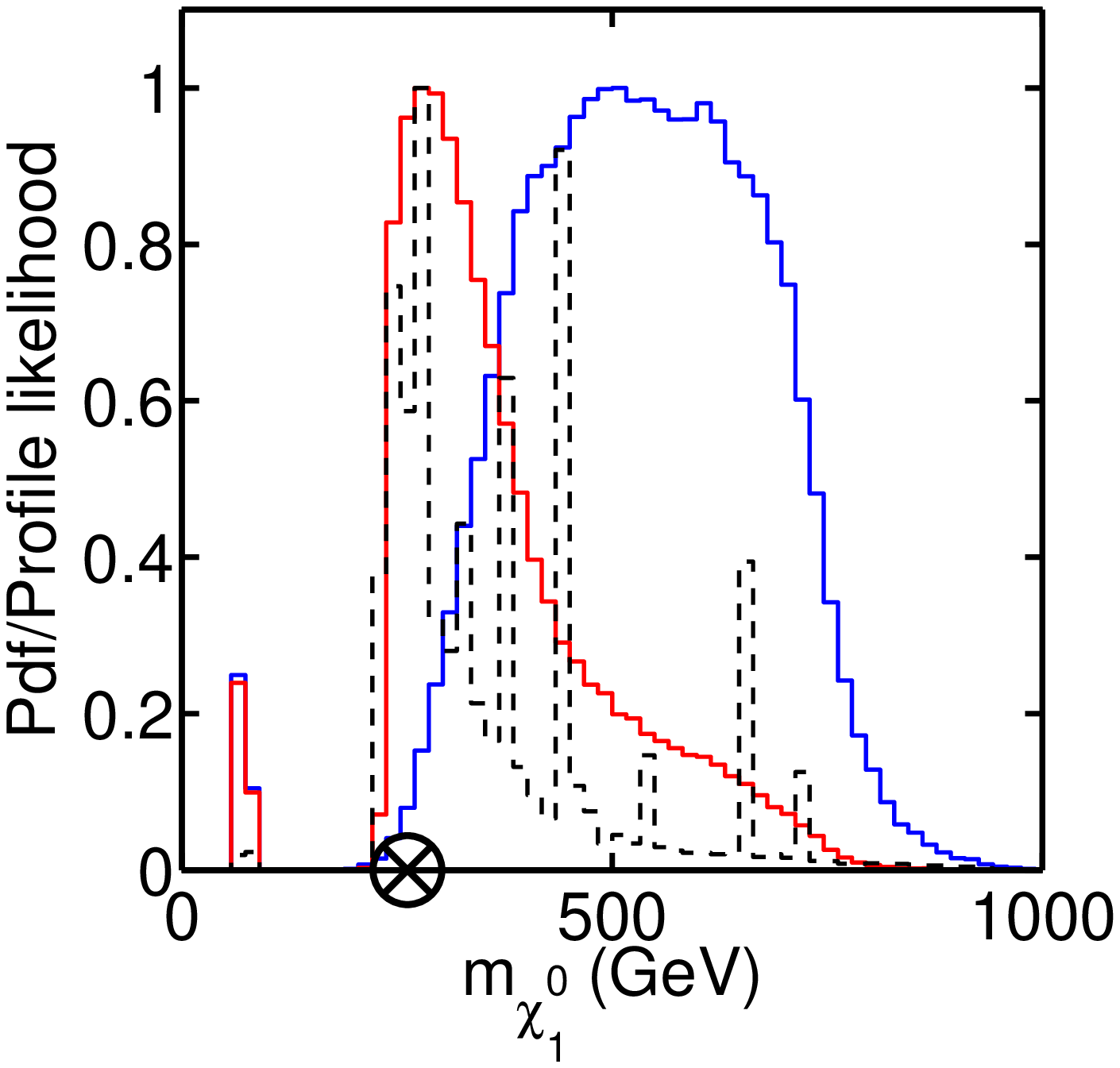}
\includegraphics[width=0.242\linewidth]{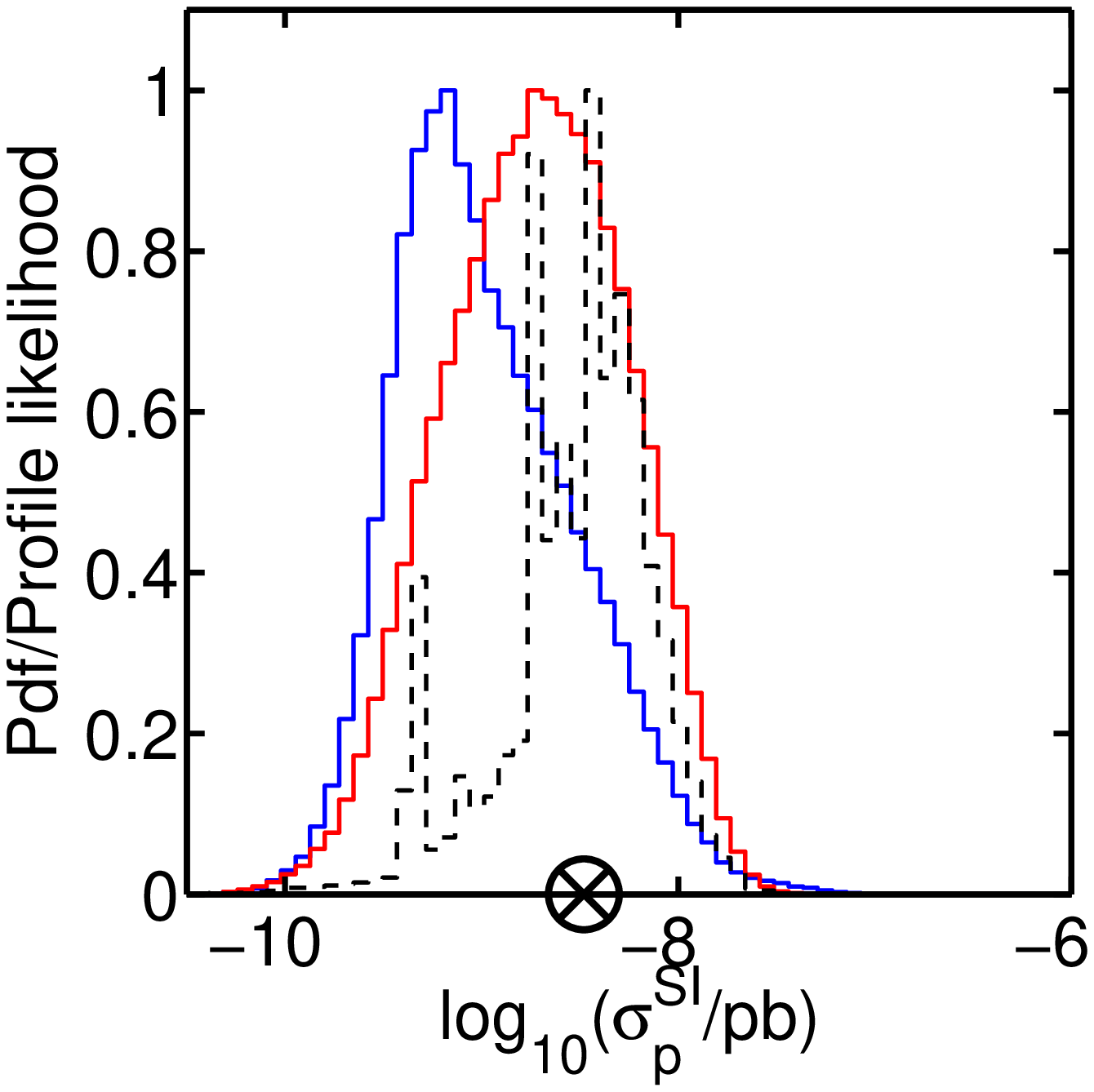}
\includegraphics[width=0.242\linewidth]{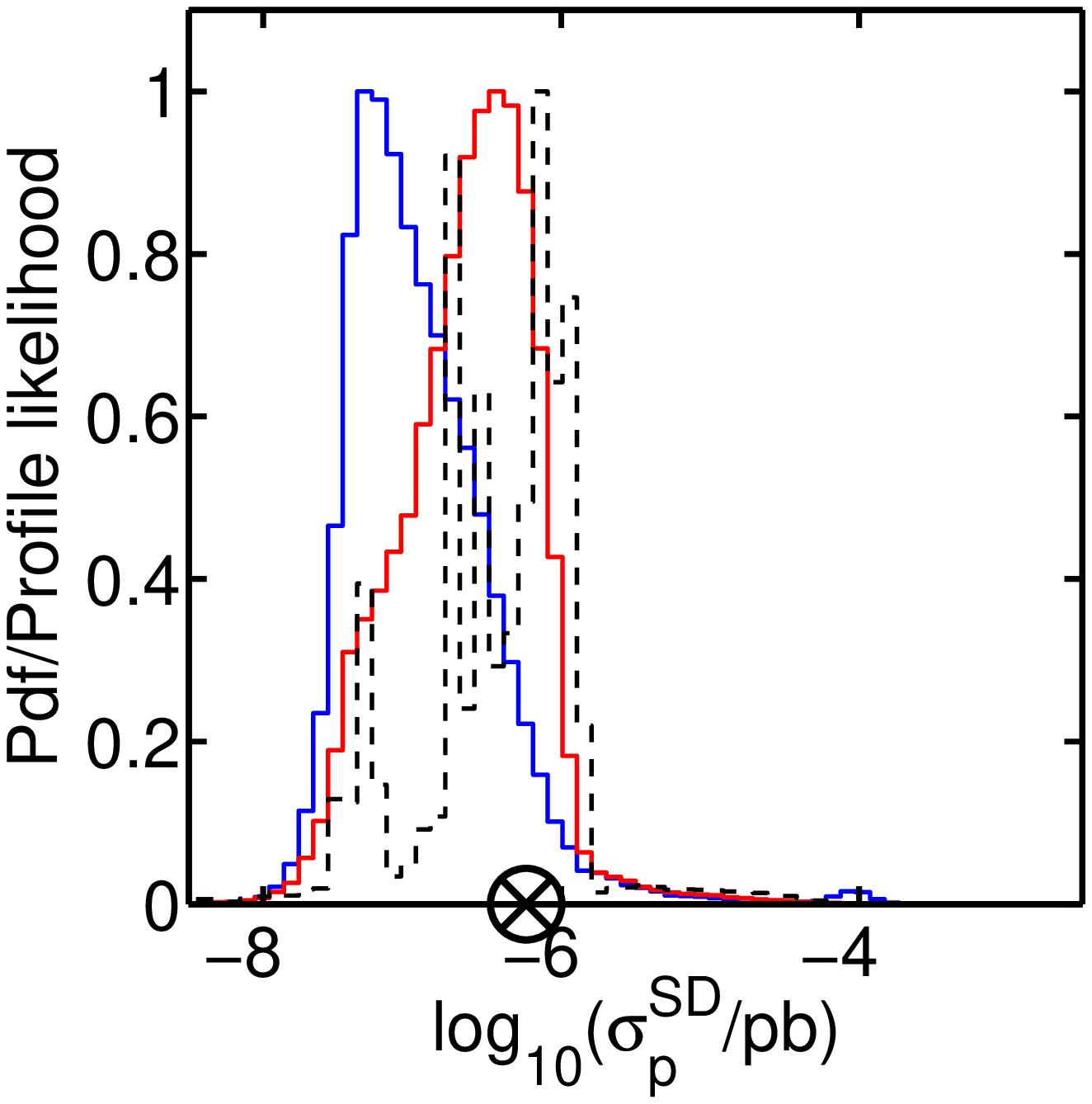}
\caption{\fontsize{9}{9} \selectfont 1D marginal pdf for flat priors (thin solid/blue), log priors (thick solid/red) and 1D profile likelihood (dashed/black) for the gluino mass, the neutralino mass, the SI and the SD scattering cross section (from left to right). Top panels include all data except XENON100, bottom panels include XENON100 data with hadronic and astrophysical uncertainties fully marginalised/maximised over. The best fit point is indicated by the encircled black cross. \label{fig:prospects}} 
\end{figure*}

We now present the implications of the above results for the prospects of detecting the cMSSM using the LHC, direct detection and indirect detection experiments. In Fig.~\ref{fig:prospects} we show the 1D marginal posterior distributions for both log (thick solid/red) and flat (thin solid/blue) priors and the 1D profile likelihood functions (dashed/black) for the gluino mass  $m_{\tilde{g}}$,  the mass of the lightest neutralino $m_{\neut}$, the spin-independent scattering cross section $\sigmaSI$ and the spin-dependent neutralino-proton scattering cross section $\sigmaSD$. 

The 1D marginal posterior distributions for the log prior scan are quite similar to the 1D profile likelihood results. The 1D posterior pdf for the flat prior scan can differ from these distributions, especially for the gluino and the neutralino masses. This is due to the flat prior suffering from residual volume effects. Comparing these results with the corresponding plots for the LHC 2010 data presented in Ref.~\cite{arXiv:1107.1715} it can be seen that now larger gluino masses $m_{\tilde{g}} \geq 1500$ GeV are favoured.
The only exception to this is the very narrow area where 60\,GeV neutralinos have a resonant annihilation through $s$-channel diagrams mediated by the lightest Higgs. The gluino mass in those cases is around 450\,GeV. 
This is apparent in the 1D plots for the gluino and neutralino masses in Fig.~\ref{fig:prospects}.

A sizable range of neutralino masses $m_{\chi} \sim 100 - 200$ GeV is excluded when combining LHC 2011 and XENON100 constraints. This includes part of the mass region which was previously favoured by the LHC 2010 results. Additionally, cross-sections $\sigmaSI$ larger than ~$10^{-8}$ pb are strongly disfavoured. However, prospects for detection of the cMSSM by the next generation of direct detection experiments remain good, with our best fit point easily within reach of the future XENON1T experiment. This experiment is expected to probe the vast majority of the current $99\%$ contours by 2015 and exclude cross-sections $\sigmaSI > 10^{-10}$ pb for WIMP masses up to $300$ GeV (see e.g. Ref.~\cite{xenon1T}). The expected 90$\%$ exclusion limit of the XENON1T experiment is indicated on the bottom right-hand panel in Fig.~\ref{fig:ALL_Xe_LHC}. The LHC 2011 limit also tightens the constraints on the spin-dependent cross-section $\sigmaSD$, strongly disfavouring $\sigmaSD > 10^{-6}$ pb for both the posterior and the profile likelihood contours. This further reduces indirect detection prospects of the cMSSM and makes its detection by the IceCube neutrino observatory, which is able to probe $\sigmaSD > 3 \times 10^{-5}$ pb (see e.g. Ref. \cite{Collaboration:2011ec}), highly unlikely.

\begin{figure*}
\centering
\includegraphics[width=0.32\linewidth]{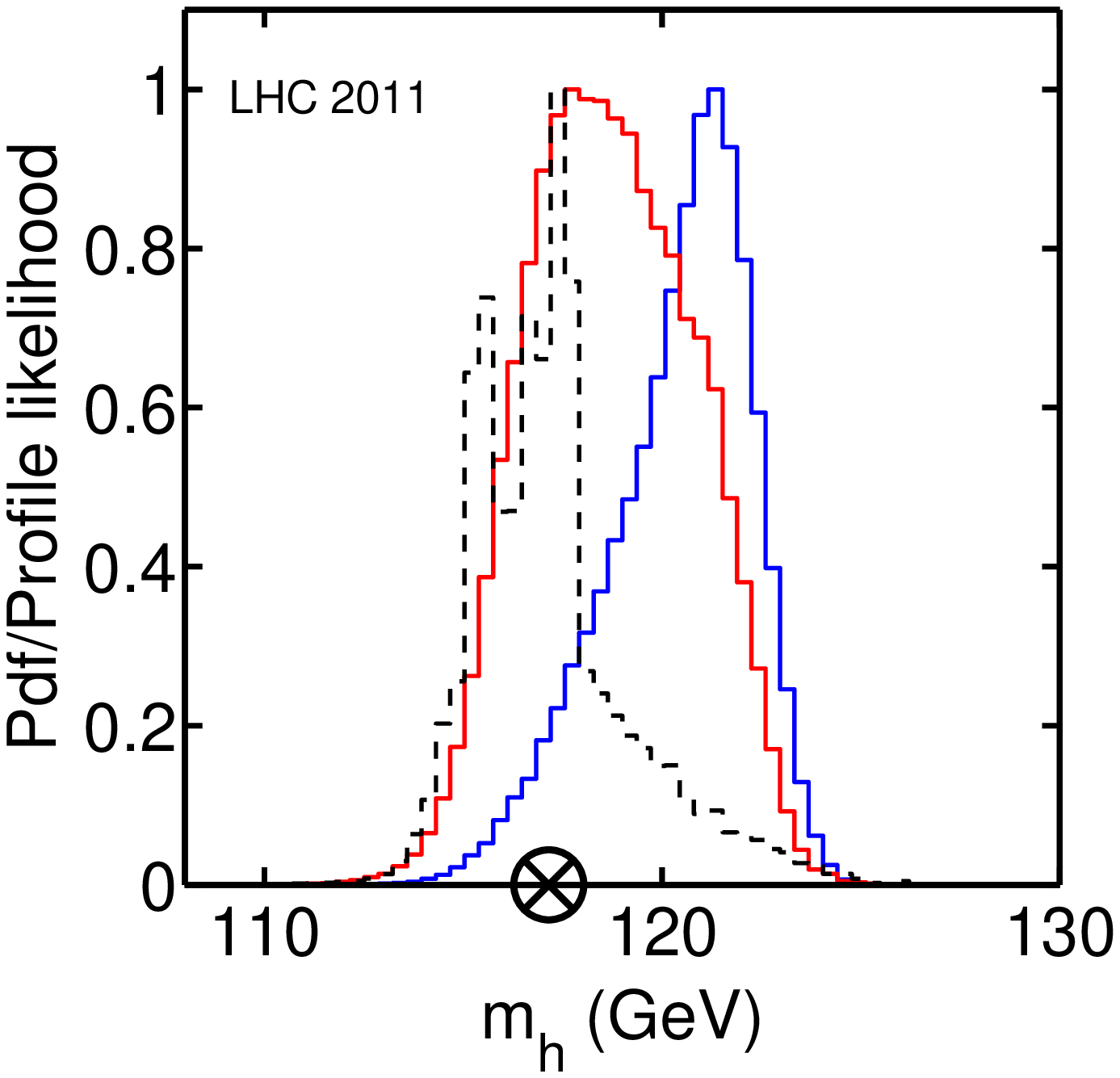}
\includegraphics[width=0.32\linewidth]{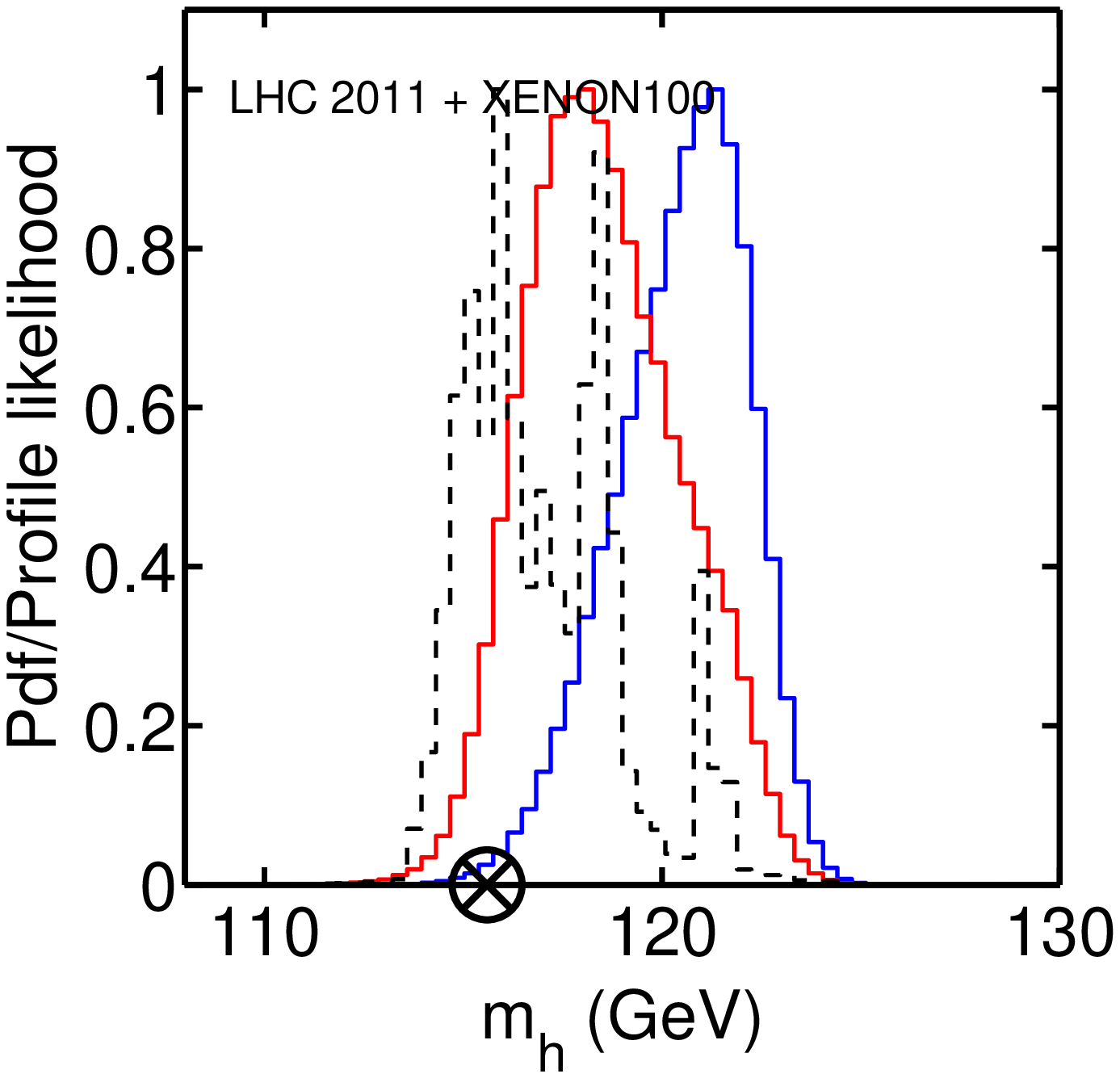}
\includegraphics[width=0.32\linewidth]{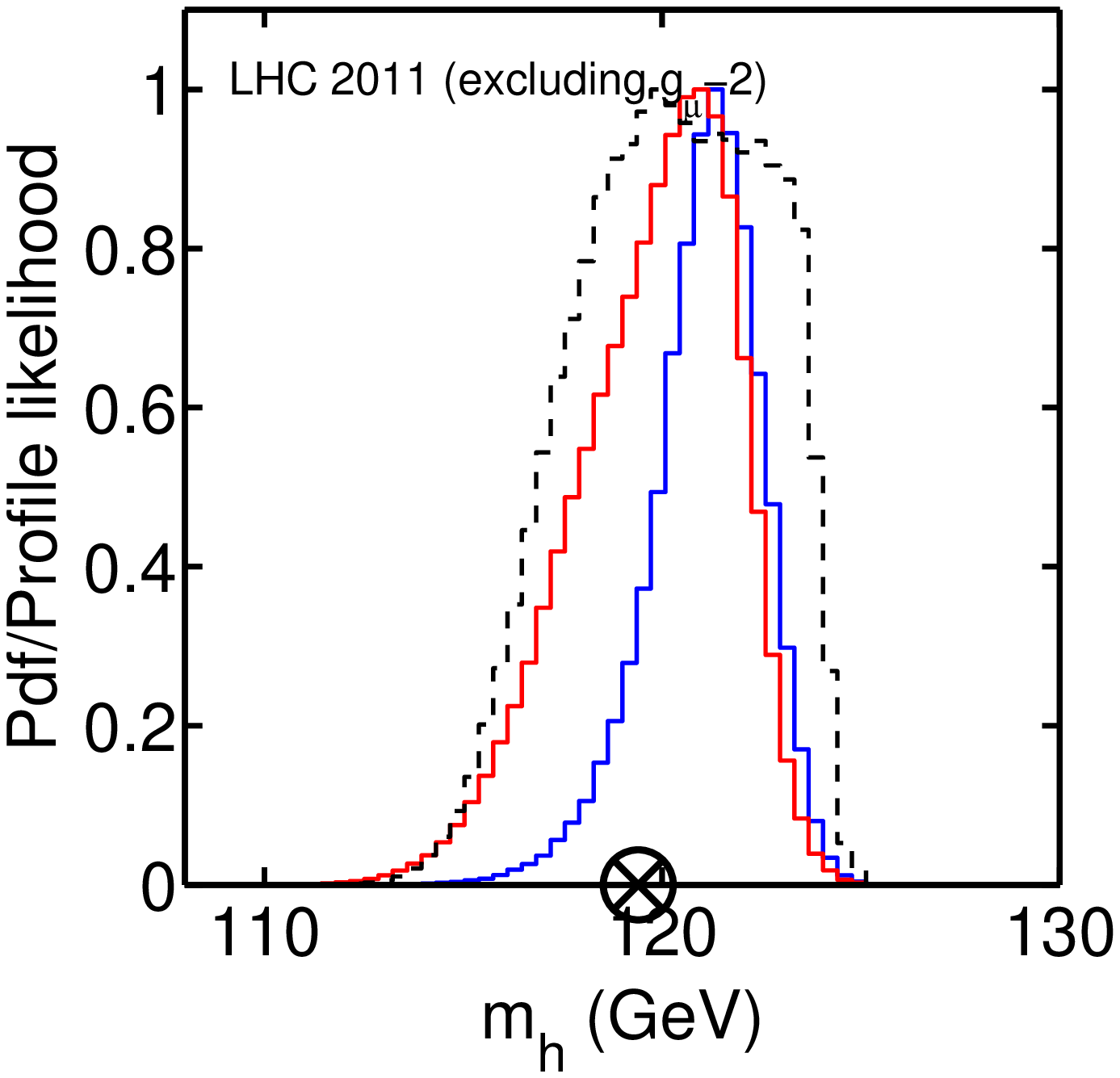}
\caption{\fontsize{9}{9} \selectfont 1D marginal pdf  for flat priors (thin solid/blue), log priors (thick solid/red) and 1D profile likelihood (dashed/black) for the lightest Higgs mass $m_h$. The results come from the implementation of all experimental data, including LHC 2011 data, except for direct detection constraints (left), all data including XENON100 data with astrophysical and hadronic uncertainties fully marginalised/maximised over (centre) and all data except direct detection data and excluding the  $\delta a_\mu^{SUSY}$ constraint. The best fit point is indicated by the encircled black cross. \label{fig:Higgs_mass}} 
\end{figure*}

The effect of LHC 2011 constraints on the mass of the lightest Higgs $m_h$ are presented in Fig.~\ref{fig:Higgs_mass}. From left to right we show constraints on $m_h$ including LHC 2011 limits but no direct detection data, LHC 2011 constraints combined with XENON100 data and LHC 2011 limits without XENON100 data and excluding the constraint on $\delta a_\mu^{SUSY}$. Compared to LHC 2010 results, small Higgs masses are now strongly disfavoured, so that the probability distributions move to larger $m_h$. This is a result of the new 95$\%$ lower limit $m_h > 115.5$ GeV. In the left and central plots the profile likelihood and the posterior pdf for log priors peak relatively closely to this new lower limit, while the flat prior scan favours slightly higher values of $m_h$. When excluding the constraint on the magnetic moment of the muon from the analysis slightly higher masses $m_h \gtrsim 120$ GeV are favoured from both the Bayesian and the profile likelihood statistical perspective (see right-hand panel in Fig.~\ref{fig:Higgs_mass}).

While we did not implement the excess at $m_h \sim 126$ GeV reported by the ATLAS collaboration explicitly, it is still obvious from Fig.~\ref{fig:Higgs_mass} that, should this excess be confirmed, the cMSSM will be strongly disfavoured. Both the frequentist profile likelihood and the Bayesian posterior probability approach zero at $m_h = 126$ GeV, independent of the choice of prior. This conclusion remains true even if the $\delta a_\mu^{SUSY}$ constraint is excluded from the scan (see right panel of Fig.~\ref{fig:Higgs_mass}).

At this point we comment on how we expect the results of our analysis to change when using priors other than the non-informative flat and log priors applied in this work. 

In the literature Bayesian studies of the cMSSM have been performed which attempt to incorporate the SUSY naturalness criterium. Namely, SUSY soft-masses should not be far from the experimental electroweak (EW) 
scale in order to avoid unnatural fine-tuning to obtain the correct size of the EW symmetry breaking. 
In some studies a penalisation of the fine-tuned regions has been implemented, e.g. by using a conveniently modified prior for the cMSSM parameters \cite{Allanach:2006jc, Allanach:2007qk}. 
On the other hand, in Ref. \cite{Cabrera:2008tj} it has been shown that the naturalness arguments arise from the Bayesian analysis itself, with no need of introducing ``naturalness priors''. 
The key is when the experimental value of $M_Z$ is considered in the same way as other experimental information (usually $M_Z$ is fixed to its experimental value and the Higgsino mass parameter $\mu$ is predicted from the EW symmetry breaking conditions). Marginalising over $\mu$ results in a factor $1/c_\mu$ in the Bayesian posterior, where $c_\mu = \left|\partial \ln M_Z^2 / \partial \ln \mu\right|$ is the conventional Barbieri-Giudice measure of the degree of fine-tuning \cite{Ellis:1986yg, Barbieri:1987fn} (for details on this derivation see Ref.~\cite{Cabrera:2008tj, Cabrera:2009dm}). 
This precisely agrees with the ``naturalness prior'' which is introduced by hand in Ref. \cite{Allanach:2006jc}. 
Thus, the presence of this fine-tuning parameter in the denominator penalises the regions of parameter space 
corresponding to large fine-tuning. As a result the only region with large soft-masses that is not disfavoured is the FP region, in which naturalness is preserved \cite{Feng:1999zg}. 
Indeed, this region contains a large portion of the Bayesian posterior probability in the presence of the DM relic abundance constraint, especially when the constraint on the anomalous magnetic moment of the muon is excluded from the analysis \cite{Cabrera:2009dm}. 
As was shown above, the addition of XENON100 data strongly disfavours the FP region, therefore one would expect the bulk of the posterior probability to fall within the low and intermediate soft-masses region, leading to similar conclusions as the ones resulting from our log prior scan.

\section{Conclusion}
\label{secconclusion}

In this work we have presented new global fits of the cMSSM, including exclusion limits from LHC 2011 SUSY and Higgs searches and XENON100 direct detection data. In addition to uncertainties on Standard Model quantities, our analysis takes into account both astrophysical and hadronic uncertainties that enter the analysis when translating direct detection limits into constraints on the cMSSM parameter space. XENON100 data in combination with LHC 2011 constraints can rule out a significant portion of the cMSSM parameter space. In addition to the FP region being ruled out at the $99\%$ level by XENON100 data, we have shown that the size of the SC region is significantly reduced by LHC 2011 constraints on $m_0$ and $m_{1/2}$. LHC 2011 data strongly impacts on the cMSSM parameter space, pushing allowed particle masses to higher values and reducing the favoured values of both spin-dependent and  spin-independent neutralino-proton cross-sections.

This has important implications for direct detection of the cMSSM, with LHC constraints cutting into the low $m_{\neut}$ region of cMSSM parameter space and, in combination with direct detection constraints, pushing the contours to lower $\sigmaSI$. Our results further emphasise the complementarity of collider experiments and direct detection searches. While LHC data are placing tight constraints on large portions of cMSSM parameter space, direct detection data is crucial to rule out regions of high SUSY masses, such as the FP region, which is unreachable for the LHC in the near future (see also the discussion in Ref. \cite{Bertone:2011kb}).

We further investigated the importance of the experimental constraint on the anomalous magnetic moment of the muon. We found that this constraint is the dominating reason for large scalar and gaugino masses being disfavoured in our analysis. All other constraints (aside from XENON100) leave this range of masses fairly unconstrained. 

Finally, we investigated constraints on the lightest Higgs mass resulting from LHC 2011 and XENON100 data. $m_h$ is pushed towards larger values due to the new LHC 2011 95$\%$ confidence level. We have qualitatively shown that, should the excess around $m_h \sim 126$ GeV be confirmed by future data sets, the residual cMSSM parameter space will be strongly disfavoured.

While this work was under completion, similar analyses of the constraints on the cMSSM parameter space resulting from LHC 2011 SUSY searches and XENON100 data were performed in Ref.~\cite{arXiv:1111.6098,arXiv:1110.3568}. Our conclusions are qualitatively similar to Ref.~\cite{arXiv:1110.3568}, while in Ref.~\cite{arXiv:1111.6098} it was claimed that the FP region can not be excluded at $95\%$ level by XENON100 data and a negative outlook on direct detection prospects of the cMSSM was given. This is a result of a very conservative estimation of the XENON100 exclusion limit in Ref.~\cite{arXiv:1111.6098}. Our analysis differs from Ref.~\cite{arXiv:1110.3568}, where astrophysical uncertainties are not included in the analysis and results are presented for the profile likelihood only. In contrast, we provide a detailed comparison between Bayesian and profile likelihood results. Ref.~\cite{arXiv:1111.6098} provides a similar comparison, however, the scanning parameters chosen are not suitable for an accurate mapping of the profile likelihood function (according to the prescriptions given in Ref.~\cite{Feroz:2011bj}), so that the extent of the resulting confidence levels may vary strongly from scan to scan. Additionally, in Ref.~\cite{arXiv:1111.6098} the impact of astrophysical and hadronic uncertainties is taken into account by smearing out the XENON100 exclusion limit, instead of marginalising/maximising over the corresponding nuisance parameters. In contrast to both Ref.~\cite{arXiv:1111.6098} and Ref.~\cite{arXiv:1110.3568} we include recent 5 fb$^{-1}$ LHC constraints on the Higgs mass, so that we provide a more sophisticated, up-to-date statistical analysis of the cMSSM parameter space.


\acknowledgments  
R.T. and C.S. would like to thank the GRAPPA Institute for hospitality. C.S. is partially supported by a scholarship of the ``Studienstiftung des deutschen Volkes''.
D.G.C. is supported by the Ram\'on y Cajal program of the Spanish MICINN and also thanks the support of the MICINN under grant FPA2009-08958, the Community of Madrid under grant HEPHACOS S2009/ESP-1473, and the European Union under the Marie Curie-ITN program PITN-GA-2009-237920. We thank the support of the Consolider-Ingenio 2010 Programme under grant MultiDark CSD2009-00064. This research was supported in part by the National Science Foundation under Grant No. NSF PHY05-51164. The use of Imperial's High Performance Computing cluster is gratefully acknowledged.
The work of G.B. is supported by the ERC Starting Grant {\it WIMPs Kairos}.


\begin{thebibliography}{99}

\bibitem{arXiv:1111.4820}
  A.~Cakir,
  arXiv:1111.4820 [hep-ph].
  
\bibitem{arXiv:1109.2352} 
  S.~Chatrchyan {\it et al.} [CMS Collaboration],
  arXiv:1109.2352 [hep-ex].

\bibitem{arXiv:1109.6572} 
  G.~Aad {\it et al.} [ATLAS Collaboration],
  arXiv:1109.6572 [hep-ex].

\bibitem{Chamseddine:1982jx}
  A.~H.~Chamseddine, R.~L.~Arnowitt and P.~Nath,
  Phys.\ Rev.\ Lett.\  {\bf 49} (1982) 970.
  
\bibitem{CTP-TAMU-24-92}
  R.~L.~Arnowitt and P.~Nath,
  Phys.\ Rev.\ Lett.\ \ {\bf 69} (1992) 725.
  
\bibitem{RAL-92-005}
  G.~G.~Ross and R.~G.~Roberts,
  Nucl.\ Phys.\ B\ {\bf 377} (1992) 571.
  
\bibitem{hep-ph/9311269}
  V.~D.~Barger, M.~S.~Berger and P.~Ohmann,
  Phys.\ Rev.\ D\ {\bf 49} (1994) 4908
  [hep-ph/9311269].

\bibitem{Kane:1993td}
  G.~L.~Kane, C.~F.~Kolda, L.~Roszkowski and J.~D.~Wells,
  Phys.\ Rev.\  D {\bf 49} (1994) 6173

\bibitem{Bertone:2010zz}
{\it Particle Dark Matter: Observations, Models and Searches}, ed. G. Bertone, Cambridge University Press (2010)

\bibitem{Bergstrom:2000pn}
  L.~Bergstrom,
  Rept.\ Prog.\ Phys.\  {\bf 63}, 793 (2000)

\bibitem{Munoz:2003gx}
  C.~Munoz,
  Int.\ J.\ Mod.\ Phys.\  A {\bf 19}, 3093 (2004)

\bibitem{Bertone:2004pz}
  G.~Bertone, D.~Hooper and J.~Silk,
  Phys.\ Rept.\  {\bf 405}, 279 (2005)
  
    \bibitem{Higgs_results}
  Available on: http://cdsweb.cern.ch/record/1406786

\bibitem{arXiv:1107.1715}
  G.~Bertone, D.~G.~Cerdeno, M.~Fornasa, R.~R.~de Austri, C.~Strege and R.~Trotta,
  arXiv:1107.1715 [hep-ph].
  
\bibitem{xenon:2011hi}
   E.~Aprile {\it et al.} 
    [XENON100 Collaboration],
  arXiv:1104.2549 [astro-ph.CO].
  
\bibitem{arXiv:1106.2529} 
  O.~Buchmueller, R.~Cavanaugh, D.~Colling, A.~De Roeck, M.~J.~Dolan, J.~R.~Ellis, H.~Flacher and S.~Heinemeyer {\it et al.},
  Eur.\ Phys.\ J.\ C\ {\bf 71}, 1722  (2011)
  [arXiv:1106.2529 [hep-ph]].
  
\bibitem{arXiv:1104.3572} 
  M.~Farina, M.~Kadastik, D.~Pappadopulo, J.~Pata, M.~Raidal and A.~Strumia,
  Nucl.\ Phys.\ B\ {\bf 853}, 607  (2011)
  [arXiv:1104.3572 [hep-ph]].
  
\bibitem{arXiv:1111.6098} 
  A.~Fowlie, A.~Kalinowski, M.~Kazana, L.~Roszkowski and Y.~L.~S.~Tsai,
  arXiv:1111.6098 [hep-ph].
  
\bibitem{arXiv:1105.5162} 
  S.~Profumo,
  Phys.\ Rev.\ D\ {\bf 84}, 015008  (2011)
  [arXiv:1105.5162 [hep-ph]].

\bibitem{arXiv:1110.3568} 
  O.~Buchmueller, R.~Cavanaugh, A.~De Roeck, M.~J.~Dolan, J.~R.~Ellis, H.~Flacher, S.~Heinemeyer and G.~Isidori {\it et al.},
  arXiv:1110.3568 [hep-ph].
  
\bibitem{Buchmueller:2011ab} 
  O.~Buchmueller, R.~Cavanaugh, A.~De Roeck, M.~J.~Dolan, J.~R.~Ellis, H.~Flacher, S.~Heinemeyer and G.~Isidori {\it et al.},
  arXiv:1112.3564 [hep-ph].
  
   \bibitem{Feroz:2007kg}
F.~Feroz and M.~P. Hobson, 
 {Mon. Not. Roy. Astron. Soc.} {\bf 384} (2008) 449--463.
  
    \bibitem{deAustri:2006pe}
  R.~Ruiz~de Austri, R.~Trotta and L.~Roszkowski,
  JHEP {\bf 0605}, 002 (2006)

\bibitem{Trotta:2008bp}
  R.~Trotta, F.~Feroz, M.~P.~Hobson, L.~Roszkowski and R.~Ruiz de Austri,
  JHEP {\bf 0812} (2008) 024
  
      \bibitem{Feroz:2011bj}
  F.~Feroz, K.~Cranmer, M.~Hobson, R.~Ruiz de Austri and R.~Trotta,
  JHEP in print, 
  arXiv:1101.3296 [hep-ph].
  
      \bibitem{ST}
  L.~E.~Strigari and R.~Trotta,
  JCAP {\bf 0911}, 019 (2009)
  
\bibitem{Feroz:2008wr} 
  F.~Feroz, B.~C.~Allanach, M.~Hobson, S.~S.~AbdusSalam, R.~Trotta and A.~M.~Weber,
  JHEP {\bf 0810}, 064 (2008)
  [arXiv:0807.4512 [hep-ph]].

  \bibitem{Ellis:2001nx}
  J.~R.~Ellis, K.~A.~Olive and Y.~Santoso,
  Astropart.\ Phys.\  {\bf 18} (2003) 395
  [hep-ph/0112113].
  
    \bibitem{Roszkowski:2007fd}
  L.~Roszkowski, R.~Ruiz de Austri and R.~Trotta,
  JHEP {\bf 0707}, 075 (2007)
  
\bibitem{Lisanti:2010qx} 
  M.~Lisanti, L.~E.~Strigari, J.~G.~Wacker and R.~H.~Wechsler,
  Phys.\ Rev.\ D {\bf 83}, 023519 (2011)
  [arXiv:1010.4300 [astro-ph.CO]].
  
\bibitem{Vogelsberger:2008qb} 
  M.~Vogelsberger, A.~Helmi, V.~Springel, S.~D.~M.~White, J.~Wang, C.~S.~Frenk, A.~Jenkins and A.~D.~Ludlow {\it et al.},
  arXiv:0812.0362 [astro-ph].
  
\bibitem{Read:2008fh} 
  J.~I.~Read, G.~Lake, O.~Agertz and V.~P.~Debattista,
  arXiv:0803.2714 [astro-ph].
  
\bibitem{Bruch:2008rx} 
  T.~Bruch, J.~Read, L.~Baudis and G.~Lake,
  Astrophys.\ J.\  {\bf 696}, 920 (2009)
  [arXiv:0804.2896 [astro-ph]].
  
\bibitem{Purcell:2009yp} 
  C.~W.~Purcell, J.~S.~Bullock and M.~Kaplinghat,
  Astrophys.\ J.\  {\bf 703}, 2275 (2009)
  [arXiv:0906.5348 [astro-ph.GA]].
  
\bibitem{Schneider:2010jr} 
  A.~Schneider, L.~Krauss and B.~Moore,
  Phys.\ Rev.\ D {\bf 82}, 063525 (2010)
  [arXiv:1004.5432 [astro-ph.GA]].
  
\bibitem{Serpico:2010ae} 
  P.~D.~Serpico and G.~Bertone,
  Phys.\ Rev.\ D {\bf 82}, 063505 (2010)
  [arXiv:1006.3268 [astro-ph.HE]].
  
    \bibitem{Trotta:2008qt}
  R.~Trotta,
  Contemp.\ Phys.\  {\bf 49}, 71 (2008)
  
      \bibitem{Scott:2009jn}
  P.~Scott, J.~Conrad, J.~Edsjo, L.~Bergstrom, C.~Farnier and Y.~Akrami,
  JCAP {\bf 1001} (2010) 031
  
\bibitem{Easson:2009kk} 
  D.~A.~Easson and R.~Gregory,
  Phys.\ Rev.\ D {\bf 80}, 083518 (2009)
  [arXiv:0902.1798 [hep-th]].
  
  \bibitem{SuperBayeS}
Available from:  \texttt{http://superbayes.org/}

\bibitem{DarkSUSY}
  P. Gondolo, J. Edsj\"o, P. Ullio, L. Bergstr\"m, M. Schelke, E.A. Baltz, T. Bringmann and G. Duda, \texttt{http://www.darksusy.org/}
  
\bibitem{Gondolo:2004sc}
  P.~Gondolo, J.~Edsjo, P.~Ullio, L.~Bergstrom, M.~Schelke and E.~A.~Baltz,
  JCAP {\bf 0407} (2004) 008

\bibitem{SoftSUSY}
  \texttt{http://projects.hepforge.org/softsusy/}

\bibitem{Allanach:2001kg}
  B.~C.~Allanach,
  Comput.\ Phys.\ Commun.\  {\bf 143} (2002) 305

\bibitem{MicrOMEGAs}
  \texttt{http://lapth.in2p3.fr/micromegas/}

\bibitem{Belanger:2006is}
  G.~Belanger, F.~Boudjema, A.~Pukhov and A.~Semenov,
  Comput.\ Phys.\ Commun.\  {\bf 176} (2007) 367

\bibitem{SuperIso}
  \texttt{http://superiso.in2p3.fr/}

\bibitem{Mahmoudi:2008tp}
F.~Mahmoudi,
Comput.\ Phys.\ Commun.\  {\bf 180}, 1579 (2009)
  
  \bibitem{Degrassi:2007kj}
  G.~Degrassi, P.~Gambino and P.~Slavich,
  Comput.\ Phys.\ Commun.\  {\bf 179} (2008) 759
  
  \bibitem{SusyBSG}
  \texttt{http://slavich.web.cern.ch/slavich/susybsg/}
  

\bibitem{Feroz:2008xx}
F.~Feroz, M.~P. Hobson, and M.~Bridges, 
  {Mon. Not. Roy. Astron. Soc.} {\bf 398} (2009) 1601--1614.
  
\bibitem{Allanach:2011ut}
  B.~C.~Allanach,
  Phys.\ Rev.\  D {\bf 83}, 095019 (2011)
  [arXiv:1102.3149 [hep-ph]].
  
\bibitem{Feng:1999zg}
  J.~L.~Feng, K.~T.~Matchev and T.~Moroi,
  Phys.\ Rev.\  D {\bf 61}, 075005 (2000)

\bibitem{Feng:2000gh}
  J.~L.~Feng, K.~T.~Matchev and F.~Wilczek,
  Phys.\ Lett.\  B {\bf 482}, 388 (2000)
    
\bibitem{Chan:1997bi}
  K.~L.~Chan, U.~Chattopadhyay and P.~Nath,
  Phys.\ Rev.\  D {\bf 58}, 096004 (1998)
    
     \bibitem{Baer:1997ai}
  H.~Baer and M.~Brhlik,
  Phys.\ Rev.\  D {\bf 57}, 567 (1998)
  
\bibitem{Akeroyd:2009tn}
 A.~G.~Akeroyd and F.~Mahmoudi,
 JHEP {\bf 0904} (2009) 121

\bibitem{Ahmady:2006yr}
 M.~R.~Ahmady and F.~Mahmoudi,
 Phys.\ Rev.\  D {\bf 75} (2007) 015007
 
 
\bibitem{g-2}
  G.~W.~Bennett {\it et al.} [ Muon G-2 Collaboration ],
  Phys.\ Rev.\  {\bf D73 } (2006)  072003.
  [hep-ex/0602035].

\bibitem{Jegerlehner:2009ry}
  F.~Jegerlehner and A.~Nyffeler,
  Phys.\ Rept.\  {\bf 477} (2009) 1
  [arXiv:0902.3360 [hep-ph]].
  
  \bibitem{Davier:2010nc}
  M.~Davier, A.~Hoecker, B.~Malaescu and Z.~Zhang,
  Eur.\ Phys.\ J.\ C {\bf 71} (2011) 1515
  [arXiv:1010.4180 [hep-ph]].
  
  \bibitem{Hagiwara:2011af}
  K.~Hagiwara, R.~Liao, A.~D.~Martin, D.~Nomura and T.~Teubner,
  J.\ Phys.\ G G {\bf 38} (2011) 085003
  [arXiv:1105.3149 [hep-ph]].

\bibitem{xenon1T}
E.~Aprile et al.  ({\small XENON1T Collaboration}), XENON1T at LNGS, Proposal, April (2010) and Technical Design Report, October (2010).

\bibitem{Collaboration:2011ec}
  I.~Collaboration,
  arXiv:1112.1840 [astro-ph.HE].
  
\bibitem{Bertone:2011kb}
  G.~Bertone, D.~Cumberbatch, R.~R.~de Austri and R.~Trotta,
  arXiv:1107.5813 [astro-ph.HE].

\bibitem{Allanach:2006jc}
  B.~C.~Allanach,
  Phys.\ Lett.\  B {\bf 635} (2006) 123
  [arXiv:hep-ph/0601089].

%
\bibitem{Allanach:2007qk}
  B.~C.~Allanach, K.~Cranmer, C.~G.~Lester and A.~M.~Weber,
  JHEP {\bf 0708} (2007) 023
  [arXiv:0705.0487 [hep-ph]].
%
\bibitem{Cabrera:2008tj}
  M.~E.~Cabrera, J.~A.~Casas and R.~Ruiz de Austri,
  JHEP {\bf 0903}, 075 (2009)
  [arXiv:0812.0536 [hep-ph]].

\bibitem{Cabrera:2009dm}
  M.~E.~Cabrera, J.~A.~Casas and R.~Ruiz d Austri,
  JHEP {\bf 1005} (2010) 043
  [arXiv:0911.4686 [hep-ph]].

\bibitem{Ellis:1986yg}
  J.~R.~Ellis, K.~Enqvist, D.~V.~Nanopoulos and F.~Zwirner,
  Mod.\ Phys.\ Lett.\  A {\bf 1} (1986) 57.

\bibitem{Barbieri:1987fn}
  R.~Barbieri and G.~F.~Giudice,
  Nucl.\ Phys.\  B {\bf 306} (1988) 63.

\bibitem{Feng:1999zg}
  J.~L.~Feng, K.~T.~Matchev and T.~Moroi,
  Phys.\ Rev.\  D {\bf 61} (2000) 075005
  [arXiv:hep-ph/9909334].



  
  %
%
%
%
%
%
%
%
%
%
%
%
%
%
%
%
%
%
%
%
%
%
%
%
%
%
%
%
%
%
%
%
%
%
%
%
%
%
%
%
%
%
%
%
%
%
%
%
%
%
%
%
%



\end{thebibliography}
\end{document}